\documentclass[letter]{aa}

\usepackage{graphicx}
\usepackage{txfonts}
\usepackage{hyperref}
\usepackage{natbib}
\usepackage{makecell}
\usepackage{xcolor}

\graphicspath{{./images/}
\bibpunct{(}{)}{;}{a}{}{,}}
\hypersetup{
    colorlinks=true,
    linkcolor=blue,
    citecolor=blue,
    filecolor=magenta,      
    urlcolor=blue,
}

\newcommand{\editone}[1]{#1}

\begin{document}

\title{The bulk metallicity of giant planets around M stars}
\titlerunning{The bulk metallicity of giant planets around M stars}

\author{
    Simon Müller\inst{1}
    \and
    Ravit Helled\inst{1}
}
\authorrunning{Müller \& Helled}

\institute{
    Department of Astrophysics, University of Zürich, \\
    Winterthurerstrasse 190, 8057 Zürich, Switzerland \\
    \email{simonandres.mueller@uzh.ch}
}

\date{Received October 1, 2024}

\abstract{
    The bulk-metallicity determination of giant exoplanets is essential to constrain their formation and evolution pathways and to compare them to the solar system. Previous studies inferred an inverse relation between the mass and bulk metallicity. However, the data almost exclusively contained planets that orbit FGK stars. The recent discoveries of giant exoplanets around M-dwarf stars present an opportunity to probe whether they follow a mass-metallicity trend different from that of their FGK counterparts.
    Using evolution models we characterised the interiors of giant exoplanets with reliable mass-radius measurements that orbit FGK and M-dwarf stars. We then inferred the mass-metallicity trends for both populations.
    We found that the bulk metallicity of giant planets around M stars is overall lower compared to those around FGK stars. \editone{This yielded mass-metallicity relations for the two populations with similar slopes but significantly different offsets. The lack of metal-rich giant planets around M dwarfs could explain the difference in the inferred offset and be a result of different formation conditions.}
    However, there were only 20 successful bulk-metallicity retrievals for the giant planets around M dwarfs, which resulted in rather large uncertainties. Therefore, it is of great importance to continue detecting these planets with both transit and radial velocities. Additionally, the characterisation of the atmospheres of giant planets around M-stars can further help to constrain their interiors and to investigate the atmosphere-interior connection. This will significantly contribute towards understanding the possible formation pathways of giant planets.
}

\keywords{planets and satellites: general, composition, gaseous planets}

\maketitle

\section{Introduction}
\label{sec:introduction}

The discovery and characterisation of exoplanets allow us to put our solar system in perspective and to study planets in more general terms. When both the planetary mass and radius are measured, information on the mean density, and therefore the planetary bulk composition can be inferred \citep[e.g.,][]{2007ApJ...669.1279S,2011ApJ...736L..29M,2014PNAS..11112622S,2019AREPS..47..141J}. Despite the degenerate nature of the problem \citep[e.g.,][]{2010ApJ...716.1208R,2014ApJ...792....1L},  various trends have been revealed when it comes to giant exoplanets. For example, we now know that the occurrence rate of giant planets increases with stellar metallicity and mass \citep{2005ApJ...622.1102F,2010PASP..122..905J,2012A&A...543A..45M,2013A&A...551A.112M,2018ApJ...860..109G,2019ApJ...873....8Z} and that there is a correlation between the mass and bulk metallicity of giant planets \citep{2016ApJ...831...64T,2023A&A...669A..24M,2024SSRv..220...61S} such that smaller planetary masses are characterised by higher planetary metallicity. Furthermore, a clear correlation between stellar and giant-planetary bulk metallicity is currently not observed \citep{2019AJ....158..239T}. Such trends (or the lack of them) can be used to constrain planet formation and evolution theory.

While the last decades led to significant progress in our understanding of giant exoplanets, it must be kept in mind that most of the available data correspond to planets that orbit sun-like stars. As a result, we still don't have a good understanding of how the observed trends and the inferred planetary properties change with stellar mass. Despite M-dwarf stars being the most common stellar type in the galaxy \citep{2015AJ....149....5W}, there are currently few observed giant planets around them. While standard planet formation theory typically predicts that giant planets are unlikely to form around such stars \citep{2008ApJ...673..502K,2022A&A...664A.180S}, several planets have been observed \citep[e.g.,][]{2013A&A...549A.134H,2015AJ....149..149B,2023AJ....165..120K,2024ApJ...962L..22D}. Although the sample is still rather limited, it can be compared to the one around FGK stars. In this work, we investigate whether there is a difference in the bulk metallicity of giant planets that orbit FGK and M-dwarf stars.  

Our paper is organised as follows. In Section \ref{sec:methods}, we describe the exoplanetary sample used in this work and the methods for analysing their bulk metallicities. The results of this analysis are in Section \ref{sec:results}, where we present the inferred composition and the relations between the planetary masses, metallicities and heavy-element masses. These results as well as potential issues of the analysis are discussed in Section \ref{sec:discussion}. Section \ref{sec:conclusions} summarises the most important conclusions from this work. Finally, more details about the exoplanetary sample, additional results, posterior distributions of the fit parameters, the influence of the age-prior choice, and an investigation of correlations are presented in the Appendix.

\section{Methods}
\label{sec:methods}

The planet sample in this work is from the updated PlanetS catalog\footnote{\url{https://dace.unige.ch/exoplanets}} \citep{2024A&A...688A..59P}, which uses data from the NASA Exoplanet Archive\footnote{\url{https://exoplanetarchive.ipac.caltech.edu}}. The catalogue was designed to provide a sample of exoplanets with reliable and robust measurements with relatively low measurement uncertainties (25\% in mass and 8\% in radius). Based on the mass range of planets to be gas giants and on the limitations of our models, our sample includes planets with masses between 0.1 and 10 $M_J$. We also focus on warm giant planets with equilibrium temperatures $T_{eq} \lesssim 1000$ K corresponding to irradiation fluxes of $I_* < 2 \times 10^8$ erg/s/cm$^2$). These planets are not strongly inflated, and therefore the characterisation of their interiors is more feasible than for hot Jupiters. We extended the sample with five recently detected giant planets around M-dwarfs: TOI-762 A b and TIC-46432937 b \citep{2024arXiv240707187H}, and TOI-6383 b, TOI-5176 A b, and TOI-6034 b (Shubham Kanodia, private communication). The data were split into two sets depending on their stellar spectral types: Planets around FGK stars, and planets around M-dwarf stars (with stellar effective temperatures smaller than 3900 K). In Appendix \ref{sec:appendix_exoplanetary_sample}, we show the mass-radius, mass-density and stellar irradiation distributions for the two populations.

Using the measurements of the planetary mass ($M_p$), radius ($R$), age ($\tau$) and irradiation ($I_*$), the bulk heavy-element mass fraction ($Z$) of giant exoplanets can be inferred with thermal evolution models. Here, we used the \texttt{planetsynth} evolution models from \citet{2021MNRAS.507.2094M}, which are based a grid of evolution models pre-calculated with the Modules for Experiments in Stellar Astrophysics code (MESA; \citet{2011ApJS..192....3P,2013ApJS..208....4P,2015ApJS..220...15P,2018ApJS..234...34P,2019ApJS..243...10P,2023ApJS..265...15J}) that was modified for giant planet evolution models \citep{2020ApJ...903..147M,2020A&A...638A.121M}. These thermal evolution models are limited to a maximum planetary age of 10 Gyr, while some of the planets considered in this work are older. To address this, we implemented a new inter- and extrapolation scheme using Piecewise Cubic Hermite Interpolating Polynomials (PCHIP) from the Python package \texttt{SciPy}. PCHIP interpolation preserves monotonicity, which matches the physical constraint that a cooling planet's radius decreases with time. For ages beyond 10 Gyr, the radius evolution was extrapolated. The radius of an old planet changes very slowly, and the extrapolation is expected to perform well. We validated this approach by comparing the extrapolated evolution with direct calculations with MESA and found that the agreement is excellent: The extrapolation error was significantly below observational uncertainties.

To infer the bulk metallicity, we employed the standard Monte Carlo approach \citep[e.g.,][]{2011ApJ...736L..29M,2016ApJ...831...64T,2023FrASS..1079000M}: First, a set of planetary parameters ($M_i$, $R_i$, $\tau_i$) was drawn by sampling the prior distributions from observations. We assumed normal distributions for the mass and radius and a uniform distribution for the stellar age between the estimated minimum and maximum. \editone{We imposed a lower age limit of 10 Myr (due to the limitations of \texttt{planetsynth}), and an upper age limit corresponding to the age of the Milky Way galaxy.} When no stellar age estimate was available we assumed that the star was between 1 and 10 Gyrs old. Thermal evolution models were then used to calculate the planet's cooling to determine $Z_i$ for which the radius from the model matches the observations: $R_{\rm{model}}(Z_i \, | \, M_i, \tau_i, I_*) = R_i$. We then calculated the normalised metallicity ($Z_i / Z_*$, where $Z_* = 0.014 \cdot 10^{\rm{[Fe/H]}}$ is the stellar metallicity) and the total heavy-element mass ($M_{z,\, i} = Z_i \cdot M_i$). Repeating these steps yielded an estimate of the posterior distribution of $Z$ and $M_z$ for any given planet.

Using the No U-Turn Sampler (NUTS) Hamiltonian Monte Carlo approach (implemented in the Python package \texttt{PyMC}), we determined the $M_p$-$Z / Z_*$ and $M_p$-$M_z$ relations using Bayesian linear regression on the logarithms of the variables:  

\begin{equation}
    \label{eq:linear}
    \log y = \beta_0 + \beta_1 \log M_p \, ,
\end{equation}

\noindent
where $M_p$ is the planet mass in $M_J$, $y = Z / Z_*$ or $M_z$ in $M_\oplus$, and $\beta_{0,\, 1}$ are the intercept and slope of the linear function. From the linear fit in log-log space, we also directly calculate the parameters of the corresponding power-law relationship 

\begin{equation}
    \label{eq:power_law}
    y = \beta_3 M_p^{\beta_1} \, ,
\end{equation}

\noindent
with $\beta_3 = 10^{\beta_0}$. The uncertainties on $\beta_3$ were propagated from the fit of $\beta_0$.

Generic weakly informative priors were assumed on the fit parameters $p(\beta_0, \beta_1) \propto \mathcal{N}(\mu = 0, \sigma = 1)$, where $\mathcal{N}$ is the normal distribution. We accounted for the uncertainties in the planetary mass by assuming a generic weakly informative prior and adding a Gaussian likelihood function using the observations for the mean and the standard deviation. To ensure a robust linear regression, we used the Student's t-distribution $\mathcal{T(\mu, \sigma, \nu})$ for the likelihood function of the dependent variables, where $\mu$, $\sigma$ and $\nu$ are location, scale and normality parameters. For $\mu$ and $\sigma$ we used the mean and standard deviation from the metallicity posteriors, which accounts for the uncertainty in the dependent variables. Rather than choosing a value for $\nu$, we added an informative gamma distribution prior $p(\nu) \propto \Gamma(\alpha = 2, \beta = 0.1)$, where $\alpha$ and $\beta$ are the shape and rate parameters. This prior is suitable since its support is between zero and infinity, and for $\nu \to \infty$ the Student's t-distribution tends towards a Gaussian distribution $\mathcal{T}(\mu, \sigma, \nu \to \infty) \to \mathcal{N}(\mu, \sigma)$ with mean $\mu$ and standard deviation $\sigma$.

\section{Results}
\label{sec:results}

\subsection{Planet selection}
\label{sec:results_planet_selection}

In this section, we present the metallicities inferred by the thermal evolution models and the fits for the $M_p$-$Z / Z_*$ and $M_p$-$M_z$ relations for warm giant planets as described in Section \ref{sec:methods}. These relations were derived for two mass ranges: The full sample with $0.1 \leq M_p (M_J) \leq 10$, and a mass-limited sample with $0.3 \leq M_p (M_J) \leq 2$. The full sample included as many planets as possible within the limitations of our models, which should yield lower uncertainties of the fit parameters. However, analyses of the mass-radius relationships \citep{2017ApJ...834...17C,2017A&A...604A..83B,2023arXiv231016733E,2024A&A...686A.296M,2024A&A...688A..59P} and planet-formation considerations \citep{2023A&A...675L...8H} suggest that giant planets may be better described as the population with masses beyond about $0.3 M_J$. For the upper mass limit in this second sample, we chose 2 $M_J$. This limit was based on the masses of the current observed giant planets around M-dwarf stars and considerations of their formation pathways \citep{2014prpl.conf..643H}. We consider this mass-limited sample to represent the class of giant planets better and therefore use it here to derive the $M_p$-$Z / Z_*$ and $M_p$-$M_z$ relations. The results for the full sample are presented in Appendix \ref{sec:appendix_full_sample}.\editone{The planets, their inferred bulk metallicites and heavy-element masses are listed in Table \ref{tab:planets}}.

\subsection{Mass-limited sample}
\label{sec:results_mass_limited_sample}

There were 49 and 15 successful metallicity retrievals for planets around FGK and M-dwarf stars in the preferred mass-limited sample. The Bayesian regression resulted in the following power-law relation for $M_p$-$Z / Z_*$:

\begin{equation}
    \label{eq:mass_metallicity_limited}
    Z / Z_* = 
    \begin{cases}
        (8.01 \pm 1.19) M^{(-0.41 \pm 0.17)} & \text{FGK stars} \, ,\\
        (1.71 \pm 0.73) M^{(-0.74 \pm 0.60)} & \text{M dwarfs} \, .
    \end{cases}
\end{equation}

\noindent
where $M_p$ is in Jupiter masses. The resulting fits for the $M_p$-$M_z$ relation are given by: 

\begin{equation}
    \label{eq:mass_mass_limited}
    M_z = 
    \begin{cases}
        (46.81 \pm 9.86) M^{(0.37 \pm 0.24)} & \text{FGK stars} \, ,\\
        (20.17 \pm 6.83) M^{(0.40 \pm 0.53)} & \text{M dwarfs} \, ,
    \end{cases}
\end{equation}

\noindent
where $M_z$ is in Earth units. All fit parameters are tabulated in Appendix \ref{sec:appendix_fit_parameters} in Tables \ref{tab:mass_metallicity} and \ref{tab:mass_mass}, and their posterior distributions from the Bayesian inference are shown in Figures \ref{fig:corner_metallicity} and \ref{fig:corner_mass}.

\begin{figure}[h]
    \includegraphics[width=\columnwidth]{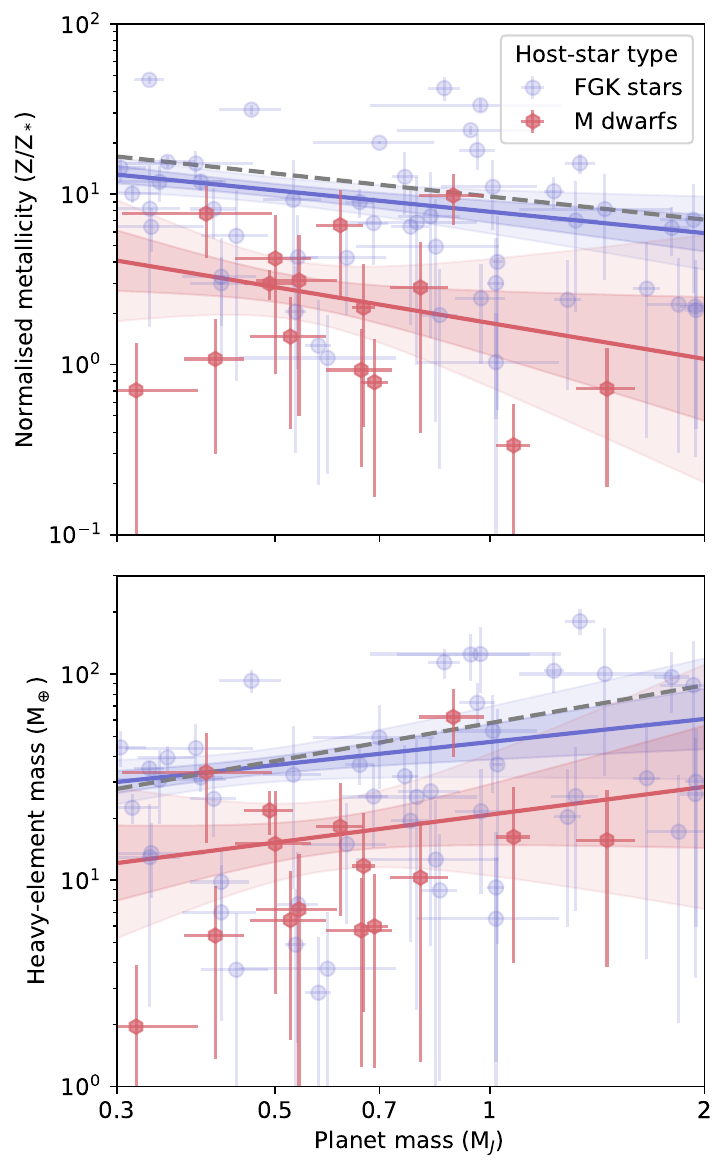}
    \caption{Normalised metallicity (top) and heavy-element mass (bottom) as a function of planet mass for $0.3 \leq M_p (M_J) \leq 2$. The scatter points and error bars show the inferred normalised metallicity or heavy element mass by the thermal evolution models. Solid lines show the fits constructed by Bayesian regression, and the shaded contours are the one and two $\sigma$ uncertainties. Planets around FGK stars are depicted in blue, while those around M-dwarfs are in red. The grey dashed line shows the fit from \citet{2016ApJ...831...64T} for comparison.}
    \label{fig:metallicity_limited}
\end{figure}

Figure \ref{fig:metallicity_limited} shows the inferred normalised metallicities, heavy-element masses, and the two fits. Both populations of planets show a decreasing trend for the normalised metallicity and an increasing trend for the heavy-element mass. There is a considerable offset between the fits for the planets around FGK and M-dwarf stars. While the scatter is fairly large, we find that giant planets around M dwarfs are overall metal-poor compared to their FGK counterparts.

We compare our results to those presented in \citet{2016ApJ...831...64T}. For both populations, the results deviate. They are quite similar for planets around FGK stars, but our results suggest a lower metal enrichment. \editone{In fact, we find numerous planets with normalised metallicities of about one or less, i.e., planets with bulk metallicities that are stellar or even sub-stellar. Similarly, we also find planets that contain less than 10 $M_\oplus$ of heavy elements, a value that traditionally corresponded to the critical core mass to initiate rapid gas accretion \citep[e.g.,][]{2014prpl.conf..643H,2018haex.bookE.140D}. Today, however, we know that the critical core mass can vary and depends on the exact formation conditions \citep[e.g.,][]{2001ApJ...553..999I,2010Icar..209..616M,2015A&A...576A.114V}.
We suggest that formation models should investigate in detail whether gas accretion is indeed initiated at smaller core mass for giant planets forming around M-dwarf stars}.
%This suggests that either the concept of a critical core mass does not apply to all giant planets, or that current thermal evolution models are predicting too small planetary radii.}

Besides the different data that was used, the strongest reason for the lower metal enrichment compared to Thorngren et al. is likely due to the hydrogen-helium equation of state: We used the \citet{2019ApJ...872...51C} equation of state, while they used SCvH \citep{1995ApJS...99..713S}. It was shown that using the Chabrier et al. equation of state leads to smaller planets due to hydrogen being denser under certain pressure-temperature conditions, leading to lower inferred metallicities \citep{2020ApJ...903..147M,2023A&A...672L...1H, 2024ApJ...971..104S,2024arXiv241021382H}. \editone{In particular, \citet{2020ApJ...903..147M} identified several planets which appear to be inflated although their irradiation is insufficiently strong to explain their observed sizes.}

The downside of the mass-limited sample was less data, and the uncertainty of the fits increased. However, as shown in Figures \ref{fig:corner_metallicity} and \ref{fig:corner_mass}, the y-intercepts for the two populations are incompatible within more than two $\sigma$. Despite the large uncertainties, we suggest that, for the currently available data, the result that giant planets around M-dwarf planets are metal-poor is robust.

\subsection{Correlations with the stellar metallicity}
\label{sec:results_correlations}

% Stellar metallicity and planetary bulk metallicity
% FKG stars
% Full range: tau = 0.14, p = 3.25e-02
% Limited range: tau = 0.27, p = 7.50e-03

% M-dwarf stars
% Full range: tau = 0.01, p = 9.74e-01
% Limited range: tau = 0.03, p = 8.69e-01

% Stellar metallicity and residual planetary bulk heavy-element mass
% FKG stars
% Full range: tau = 0.25, p = 2.06e-04
% Limited range: tau = 0.26, p = 8.31e-03

% M-stars
% Full range: tau = 0.20, p = 2.29e-01
% Limited range: tau = 0.08, p = 7.00e-01

\editone{In this section, we test whether our predictions for the planetary bulk metallicity and heavy-element mass were correlated with the stellar metallicity. Additional correlations with other stellar and planetary parameters are investigated in Appendix \ref{sec:appendix_correlations}. As in the previous analysis, we split the data into planets around FGK and M-dwarf stars and used two different mass ranges. For each pairing, we calculated Kendall's tau rank correlation coefficient and its associated p-value to determine whether the correlation is statistically significant.}

\editone{For the heavy-element mass, we used the residuals to investigate the correlation. The residual heavy-element mass of a planet was calculated with its inferred heavy-element mass divided by the value predicted Eqs. \ref{eq:mass_mass_limited} or \ref{eq:mass_mass_full} depending on whether the full or mass-limited data were used. The residual represents the deviation from the value expected from the planetary mass alone. Therefore, using the residual has the advantage of highlighting whether the deviation from the fit correlates with the stellar metallicity.}

\begin{figure}
    \centering
    \includegraphics[width=\columnwidth]{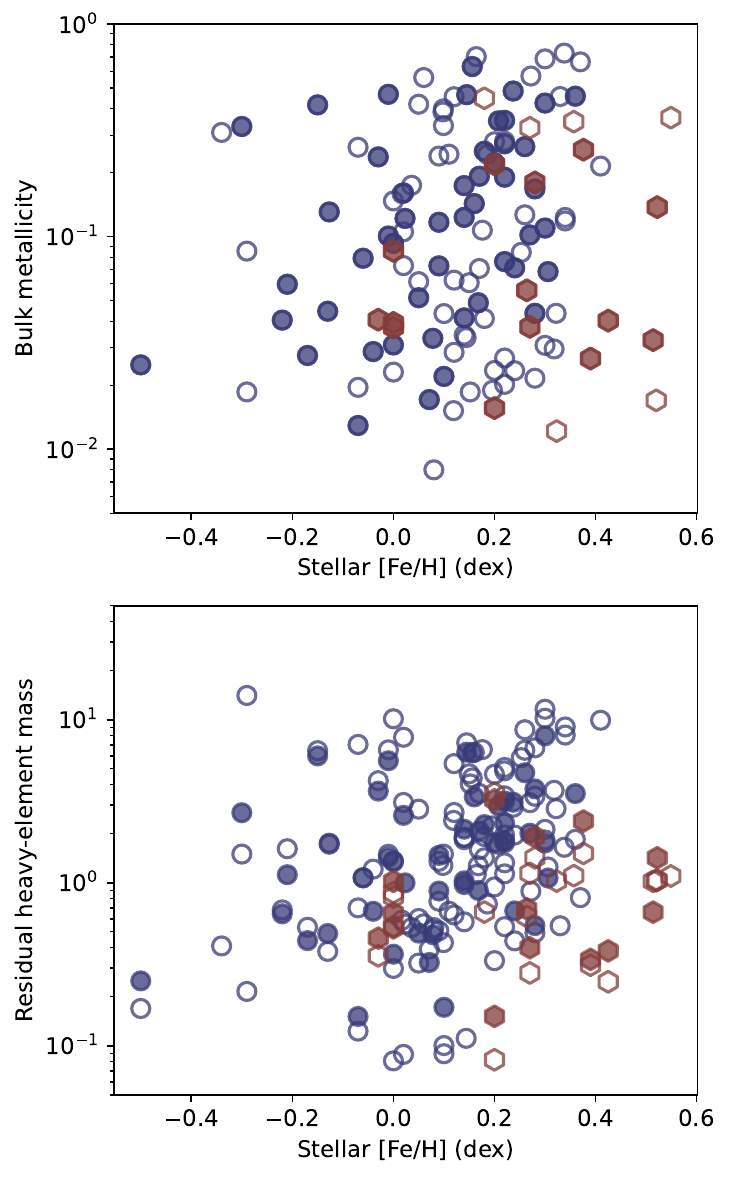}
    \caption{Top: Planetary bulk metallicity vs. stellar metallicity. Bottom: Planetary residual heavy-element mass (the ratio of the calculated and predicted heavy-element mass) vs. stellar metallicity. Planets around FGK and M-dwarf stars are depicted in blue circles and red hexagons. Filled symbols mark planets that were included in the mass-limited sample. We find that for FGK stars, the bulk metallicity and residual heavy-element mass are moderately correlated with the stellar metallicity.}
    \label{fig:correlations_st_met}
\end{figure}

\editone{The bulk metallicity and residual heavy-element mass as a function of stellar metallicity is shown in Figure \ref{fig:correlations_st_met}. For FGK stars, we find that the planetary bulk metallicity is correlated with the stellar metallicity in the full sample ($\tau = 0.14$, $p = 0.03$) and the mass-limited sample ($\tau = 0.27$, $p = 0.008$). Similarly, the stellar metallicity is also correlated with the residual heavy-element mass in both the full ($\tau = 0.25$, $p = 0.002$) and the mass-limited sample ($\tau = 0.26$, $p = 0.008$). We did not find any statistically significant correlation for the planets around M-dwarf stars. This is probably due to the limited number of detected planets.}

\editone{The inferred correlation between the stellar metallicity and the residual heavy-element mass is at odds with previous analyses \citep{2016ApJ...831...64T,2019AJ....158..239T}. In particular, \citet{2019AJ....158..239T} did not find a clear correlation between those two values, although such a correlation is predicted from formation models \citep[e.g.][]{2014A&A...572A.118M}. The different results are likely due to the significantly higher number of planets in our data. Our analysis included up to 104 planets, while \citet{2019AJ....158..239T} was limited to 24 planets. We also note that the lack of correlation for M dwarfs may be due to the limited data available. While outside the scope of this work, a thorough analysis of potential correlations between the composition of stars and their planets is desirable.}

\section{Discussion}
\label{sec:discussion}

The results presented here provide new insights into our understanding of giant planets across stellar masses. However, this is only the first step. Below, we briefly describe the limitations in terms of the data and the modelling approach as well as the connection of the results to giant planet formation theory. 

First, it should be noted that this study used a relatively low number of observed giant planets around M dwarfs, caused in part by the low occurrence rate of giant planets around M dwarfs \citep{2006ApJ...649..436E,2010PASP..122..905J,2019A&A...624A..94M,2022A&A...664A.180S,2023AJ....165...17G,2023MNRAS.521.3663B}. This leads to rather large uncertainties in the fit parameters (see Section \ref{sec:results_mass_limited_sample} and Appendix \ref{sec:appendix_full_sample}). Despite this, irrespective of the imposed mass limits the fits to the two populations were inconsistent within at least about two $\sigma$.  
There could be an observational bias towards larger, metal-poor giant planets that are more readily detected. However, measurements of low-mass and radius planets observations and higher massive planets with similar radii should not be more difficult to detect. They may, however, be rare and therefore not present in the current sample. Alternatively, due to the limited number of planets in the M-dwarf sample, it is possible that more metal-poor giant planets were observed by chance. As a result, it is desirable to have a large data sample and we therefore encourage observers to detect more giant planets around M dwarfs with both transit and radial velocity methods.

Second, the evolution models we used to infer the planetary composition are based on several assumptions and include uncertainties that could affect the results. The largest effects are from the equation of state, the assumed chemical composition of the heavy elements, the interior temperature and composition profiles, and the atmospheric model \citep{2016ApJ...831...64T,2019Atmos..10..664P,2020ApJ...903..147M,2024MNRAS.529.2242P}. However, these model uncertainties should apply to planets around FGK and M-dwarf stars equally. The two populations are fairly similar concerning their incident stellar irradiation (see Figure \ref{fig:stellar_irradiation}). As a result, differences in their atmospheres are unlikely to be caused by the heat they receive from their stars. While the absolute values of the inferred bulk metallicities would be affected by changing the model details and the inferred composition could be shifted,  the difference between the populations should remain. 

Third, a key challenge regarding the characterisation of giant planets around M-dwarfs is that the stellar ages are typically poorly constrained. When an age estimate was unavailable, we used an uninformative prior from 1 to 10 Gyr. Given that many of the M dwarfs in the sample are likely older than a few Gyr, our generous choice biases our results towards higher metallicities. This is because, for a given radius, a younger planetary age increases the inferred metallicity (compared to older ages). As a result,  our conclusion that the metallicities of giant planets around M-stars are lower than those orbiting sun-like stars is robust. Further details can be found in Appendix \ref{sec:appendix_age_prior_test} where we repeat our analysis assuming different age ranges.  

Finally, it is interesting to speculate how our results can be used to constrain giant planet formation around M dwarfs.  In the core accretion scenario \citep[e.g.,][]{1996Icar..124...62P}, the formation of a giant planet around a low-mass star takes longer mainly due to the longer orbital timescales (for a given location) and lower protoplanetary disk masses \citep{2013ApJ...771..129A,2016ApJ...831..125P}. Core-accretion planet formation models generally fail to produce giant planets within reasonable disk lifetimes \citep{2004ApJ...612L..73L,2021A&A...656A..72B,2022A&A...664A.180S}. An alternative formation pathway would be gravitational (disk) instability, in which the giant planet is formed during the proto-stellar phase of massive, gravitationally unstable protoplanetary disks \citep[e.g.,][]{1997Sci...276.1836B,2006ApJ...643..501B,2023ApJ...956....4B}.

The slopes of the mass-metallicity and $M$-$M_z$ trends we inferred here appear similar for both planet populations. These inverse relations were previously suggested to be compatible with planets that formed by core accretion \citep{2011ApJ...736L..29M,2016ApJ...831...64T}. Indeed, in the core accretion framework, the lower heavy-element content of giant planets around M dwarfs could be seen as a result of long formation timescales or less available solids. Interestingly, the slow growth of the M-dwarf giant planets is expected to lead to very extended composition gradients and non-adiabatic cooling for a large portion of the interior \citep[e.g.,][]{2017ApJ...840L...4H,2020A&A...633A..50V}. In that case, the planetary interiors could be hotter than our homogeneous evolution models predict, which would cause us to underestimate the bulk metallicity (for a given radius, a hotter planet can contain more heavy elements). 

At the same time, lower metallicities can also be a result of formation by disk instability if less heavy elements are accreted post-formation due to different disk properties and formation location \citep[e.g.,][]{2009ApJ...697.1256H}. At the moment, given the limited data and uncertainties in theoretical models, it is not possible to discriminate between these two formation paths. It is clear that more theoretical investigation of the expected metallicities of giant planets in both formation models is required.

\section{Conclusions}
\label{sec:conclusions}

We investigated the correlation between the bulk metallicity and planetary mass for giant planets around M-stars. Our exoplanetary data was mostly from the updated PlanetS catalogue but was enlarged with recently confirmed observations of giant planets around M-dwarf stars. The sample resulted in \editone{104} successful bulk-metallicity retrievals for FGK and 20 for M-dwarf stars. We performed Bayesian regression with these results to derive updated $M$-$Z / Z_*$ and $M$-$M_z$ relations. Our main results are:

\begin{itemize}
    \item The currently available data suggest that \editone{there is a lack of metal-rich giant planets around M dwarfs compared to their FGK-stars counterparts.}
    \item The $M$-$Z / Z_*$ and $M$-$M_z$ relations depend on the planetary mass range used for the fit.
    \item For masses between 0.1 and 10 M$_J$ we find the slopes and magnitudes of the metallicity and heavy-element mass fits are inconsistent between the two populations.
    \item For masses between 0.3 and 1.0, the slopes are fairly similar, but there still is a significant offset that predicts a lower bulk metallicity for giant planets around M dwarfs.
    \item \editone{For giant planets around FGK stars we find that the planetary bulk metallicity and residual heavy-element mass are moderately correlated with the stellar metallicity.}
\end{itemize}

Currently, the number of giant planets with measured mass and radius around M dwarfs is still limited. Future measurements of planets with masses between around 0.3 to 2 M$_J$ around M dwarfs will help to identify the difference between the two populations. Also, measurements of the atmospheric composition of giant planets around M dwarfs will facilitate the direct comparison of the atmospheric composition of planets around FGK and M-dwarf stars, and reveal the relationship between planetary mass and atmospheric metallicity \citep{2019ApJ...887L..20W,2023ApJS..269...31E,2024AJ....167..167S,2024AJ....167..161K}. Knowledge of the atmospheric composition, together with measured masses and radii will also improve the interior characterisation. This provides a unique opportunity to study the atmosphere-interior connection of giant planets across different stellar masses and to improve our understanding of giant planet formation in general.

\begin{acknowledgements}
    We thank Shubham Kanodia for sharing data and fruitful discussions. We also thank the anonymous reviewer for providing useful comments. We acknowledge support from the Swiss National Science Foundation (SNSF) grant \texttt{\detokenize{200020_215634}} and the National Centre for Competence in Research ‘PlanetS’ supported by SNSF. This research used data from the NASA Exoplanet Archive, which is operated by the California Institute of Technology, under contract with the National Aeronautics and Space Administration under the Exoplanet Exploration Program. 
    Extensive use was also made of the Python packages \texttt{Jupyter} \citep{jupyter}, \texttt{Matplotlib} \citep{Hunter2007}, \texttt{NumPy} \citep{harris2020array}, \texttt{pandas} \citep{mckinney-proc-scipy-2010}, \texttt{planetsynth} \citep{2021MNRAS.507.2094M}, \texttt{PyMC} \citep{pymc2023}, and \texttt{SciPy} \citep{2020SciPy-NMeth}.
\end{acknowledgements}

\bibliographystyle{aa}
\bibliography{library}

\onecolumn
\begin{center}
\tiny
\begin{longtable}{llccccccc}
\caption{\label{tab:planets} Planets in the sample and some of their properties. To improve readability, the mass and radius uncertainties are the mean of the upper and lower observational errors. For planets with missing heavy-element masses and bulk metallicites we were unable to find solutions with our evolution models. An extended version of the table is available in machine-readable format.}\\
\hline\hline
\# & Name & $M_p$ (M$_J$)& $R_p$ (R$_J$) & Flux (erg/s/cm$^2$) & Age (Gyr) & [Fe/H] (dex) & $M_z$ (M$_\oplus$) & $Z / Z_*$ \\
\hline
\endfirsthead
\caption{continued.}\\
\hline\hline
\# & Name & $M_p$ (M$_J$)& $R_p$ (R$_J$) & Flux (erg/s/cm$^2$) & Age (Gyr) & [Fe/H] (dex) & $M_z$ (M$_\oplus$) & $Z / Z_*$ \\
\hline
\endhead
\hline
\endfoot
1 & CoRoT-10 b & 2.73 $\pm$ 0.14 & 0.97 $\pm$ 0.07 & 4.56 $\times$ 10$^{7}$ & 0.01 - 3.00 & 0.26 & 109.90 $\pm$ 44.76 & 4.97 $\pm$ 2.01 \\
2 & CoRoT-20 b & 4.23 $\pm$ 0.26 & 0.84 $\pm$ 0.04 & 1.88 $\times$ 10$^{8}$ & 0.06 - 0.90 & 0.14 & - & - \\
3 & CoRoT-30 b & 2.90 $\pm$ 0.22 & 1.01 $\pm$ 0.08 & 1.45 $\times$ 10$^{8}$ & 0.60 - 6.70 & 0.02 & 97.14 $\pm$ 51.43 & 7.19 $\pm$ 3.76 \\
4 & CoRoT-8 b & 0.22 $\pm$ 0.03 & 0.57 $\pm$ 0.02 & 1.22 $\times$ 10$^{8}$ & - & 0.30 & 48.07 $\pm$ 7.53 & 24.56 $\pm$ 1.18 \\
5 & CoRoT-9 b & 0.83 $\pm$ 0.08 & 1.04 $\pm$ 0.08 & 6.69 $\times$ 10$^{6}$ & 3.00 - 9.00 & -0.01 & 26.76 $\pm$ 22.10 & 7.36 $\pm$ 5.98 \\
6 & HAT-P-12 b & 0.21 $\pm$ 0.01 & 0.96 $\pm$ 0.02 & 1.90 $\times$ 10$^{8}$ & 0.50 - 4.50 & -0.29 & 5.73 $\pm$ 3.62 & 11.90 $\pm$ 7.48 \\
7 & HAT-P-15 b & 1.94 $\pm$ 0.30 & 1.06 $\pm$ 0.07 & 1.46 $\times$ 10$^{8}$ & 4.90 - 9.30 & 0.22 & 47.70 $\pm$ 33.96 & 3.28 $\pm$ 2.25 \\
8 & HAT-P-17 b & 0.58 $\pm$ 0.06 & 1.05 $\pm$ 0.04 & 8.18 $\times$ 10$^{7}$ & 4.50 - 11.10 & 0.00 & 5.92 $\pm$ 5.44 & 2.21 $\pm$ 1.99 \\
9 & HAT-P-18 b & 0.20 $\pm$ 0.01 & 1.00 $\pm$ 0.05 & 1.17 $\times$ 10$^{8}$ & 6.00 - 16.80 & 0.10 & 2.72 $\pm$ 2.41 & 2.47 $\pm$ 2.17 \\
10 & HAT-P-20 b & 7.22 $\pm$ 0.36 & 1.02 $\pm$ 0.05 & 1.96 $\times$ 10$^{8}$ & 2.90 - 12.40 & 0.22 & 46.38 $\pm$ 26.53 & 0.87 $\pm$ 0.49 \\
11 & HAT-P-54 b & 0.76 $\pm$ 0.03 & 0.94 $\pm$ 0.03 & 1.02 $\times$ 10$^{8}$ & 1.80 - 8.20 & -0.13 & 31.61 $\pm$ 13.02 & 12.50 $\pm$ 5.09 \\
12 & HATS-17 b & 1.34 $\pm$ 0.07 & 0.78 $\pm$ 0.06 & 9.91 $\times$ 10$^{7}$ & 0.80 - 3.40 & 0.30 & 180.55 $\pm$ 26.11 & 15.21 $\pm$ 2.06 \\
13 & HATS-22 b & 2.74 $\pm$ 0.11 & 0.95 $\pm$ 0.04 & 1.22 $\times$ 10$^{8}$ & - & 0.00 & 128.16 $\pm$ 34.29 & 10.52 $\pm$ 2.78 \\
14 & HATS-47 b & 0.37 $\pm$ 0.03 & 1.12 $\pm$ 0.01 & 1.21 $\times$ 10$^{8}$ & 3.80 - 11.00 & -0.11 & - & - \\
15 & HATS-48 A b & 0.24 $\pm$ 0.03 & 0.80 $\pm$ 0.01 & 1.91 $\times$ 10$^{8}$ & 11.36 - 12.39 & 0.19 & 18.90 $\pm$ 3.48 & 11.35 $\pm$ 1.43 \\
16 & HATS-49 b & 0.35 $\pm$ 0.03 & 0.77 $\pm$ 0.01 & 1.11 $\times$ 10$^{8}$ & 8.50 - 11.90 & 0.21 & 39.43 $\pm$ 5.55 & 15.51 $\pm$ 1.25 \\
17 & HATS-6 b & 0.32 $\pm$ 0.07 & 1.00 $\pm$ 0.02 & 5.84 $\times$ 10$^{7}$ & 3.80 - 12.40 & 0.20 & 1.94 $\pm$ 1.90 & 0.70 $\pm$ 0.64 \\
18 & HATS-72 b & 0.13 $\pm$ 0.00 & 0.72 $\pm$ 0.00 & 6.75 $\times$ 10$^{7}$ & 11.72 - 12.41 & 0.10 & 13.27 $\pm$ 0.51 & 18.93 $\pm$ 0.30 \\
19 & HATS-74 A b & 1.46 $\pm$ 0.14 & 1.03 $\pm$ 0.02 & 1.48 $\times$ 10$^{8}$ & 5.90 - 16.10 & 0.51 & 15.50 $\pm$ 11.76 & 0.71 $\pm$ 0.53 \\
20 & HATS-75 b & 0.49 $\pm$ 0.04 & 0.88 $\pm$ 0.01 & 7.93 $\times$ 10$^{7}$ & 10.60 - 18.20 & 0.52 & 21.60 $\pm$ 5.13 & 2.95 $\pm$ 0.59 \\
21 & HATS-76 b & 2.63 $\pm$ 0.09 & 1.08 $\pm$ 0.03 & 1.79 $\times$ 10$^{8}$ & 0.60 - 13.30 & 0.32 & 36.44 $\pm$ 28.53 & 1.48 $\pm$ 1.16 \\
22 & HATS-77 b & 1.37 $\pm$ 0.09 & 1.17 $\pm$ 0.02 & 1.5 $\times$ 10$^{8}$ & 7.10 - 17.10 & 0.25 & - & - \\
23 & HD 114082 b & 8.00 $\pm$ 1.00 & 1.00 $\pm$ 0.03 & 2.04 $\times$ 10$^{7}$ & 0.01 - 0.02 & 0.00 & - & - \\
24 & HD 17156 b & 3.19 $\pm$ 0.03 & 1.09 $\pm$ 0.02 & 1.45 $\times$ 10$^{8}$ & 2.91 - 3.58 & 0.24 & 23.73 $\pm$ 18.10 & 0.96 $\pm$ 0.73 \\
25 & HD 332231 b & 0.24 $\pm$ 0.02 & 0.87 $\pm$ 0.03 & 1.34 $\times$ 10$^{8}$ & 2.40 - 6.80 & 0.04 & 13.56 $\pm$ 4.24 & 11.47 $\pm$ 3.37 \\
26 & HD 80606 b & 3.94 $\pm$ 0.11 & 0.98 $\pm$ 0.03 & 5.12 $\times$ 10$^{6}$ & 3.90 - 7.50 & - & 173.45 $\pm$ 48.27 & - \\
27 & HD 95338 b & 0.13 $\pm$ 0.01 & 0.35 $\pm$ 0.02 & 9.98 $\times$ 10$^{6}$ & 2.57 - 7.59 & 0.04 & - & - \\
28 & K2-114 b & 2.01 $\pm$ 0.12 & 0.93 $\pm$ 0.03 & 5.43 $\times$ 10$^{7}$ & 2.70 - 11.50 & 0.41 & 137.32 $\pm$ 37.22 & 5.97 $\pm$ 1.56 \\
29 & K2-115 b & 0.84 $\pm$ 0.19 & 1.11 $\pm$ 0.06 & 4.86 $\times$ 10$^{7}$ & 6.50 - 12.40 & -0.22 & 12.14 $\pm$ 11.81 & 4.80 $\pm$ 4.44 \\
30 & K2-139 b & 0.39 $\pm$ 0.08 & 0.81 $\pm$ 0.03 & 2.30 $\times$ 10$^{7}$ & 1.50 - 2.10 & 0.22 & 43.79 $\pm$ 13.36 & 15.16 $\pm$ 2.72 \\
31 & K2-19 b & 0.10 $\pm$ 0.01 & 0.62 $\pm$ 0.02 & 1.19 $\times$ 10$^{8}$ & - & 0.06 & 18.39 $\pm$ 1.39 & 34.87 $\pm$ 2.27 \\
32 & K2-280 b & 0.12 $\pm$ 0.02 & 0.67 $\pm$ 0.04 & 8.61 $\times$ 10$^{7}$ & 7.26 - 10.66 & 0.33 & 17.20 $\pm$ 3.59 & 15.24 $\pm$ 2.34 \\
33 & K2-287 b & 0.32 $\pm$ 0.03 & 0.85 $\pm$ 0.01 & 1.01 $\times$ 10$^{8}$ & 3.50 - 5.50 & 0.20 & 22.35 $\pm$ 3.69 & 10.03 $\pm$ 1.20 \\
34 & K2-290 c & 0.77 $\pm$ 0.05 & 1.01 $\pm$ 0.05 & 4.74 $\times$ 10$^{7}$ & 3.20 - 5.60 & -0.06 & 19.56 $\pm$ 14.47 & 6.49 $\pm$ 4.75 \\
35 & K2-295 b & 0.34 $\pm$ 0.06 & 0.90 $\pm$ 0.01 & 1.16 $\times$ 10$^{8}$ & - & 0.14 & 13.39 $\pm$ 5.32 & 6.37 $\pm$ 1.85 \\
36 & K2-329 b & 0.26 $\pm$ 0.02 & 0.77 $\pm$ 0.02 & 6.20 $\times$ 10$^{7}$ & 0.50 - 4.00 & 0.10 & 31.95 $\pm$ 5.38 & 22.01 $\pm$ 3.09 \\
37 & K2-55 b & 0.14 $\pm$ 0.02 & 0.39 $\pm$ 0.03 & 1.78 $\times$ 10$^{8}$ & - & 0.38 & - & - \\
38 & KOI-1257 b & 1.45 $\pm$ 0.35 & 0.94 $\pm$ 0.12 & 9.93 $\times$ 10$^{6}$ & 6.30 - 12.30 & 0.27 & 47.13 $\pm$ 29.51 & 3.91 $\pm$ 2.18 \\
39 & KOI-1783.01 & 0.22 $\pm$ 0.03 & 0.79 $\pm$ 0.02 & 7.03 $\times$ 10$^{6}$ & - & 0.11 & 17.51 $\pm$ 5.19 & 13.54 $\pm$ 3.03 \\
40 & KOI-3680 b & 1.93 $\pm$ 0.20 & 0.99 $\pm$ 0.07 & 4.59 $\times$ 10$^{6}$ & 0.90 - 12.80 & 0.16 & 88.23 $\pm$ 57.22 & 7.07 $\pm$ 4.48 \\
41 & KOI-94 d & 0.33 $\pm$ 0.03 & 1.01 $\pm$ 0.09 & 1.45 $\times$ 10$^{8}$ & 2.77 - 3.55 & 0.02 & 13.00 $\pm$ 10.52 & 8.25 $\pm$ 6.56 \\
42 & Kepler-111 c & 0.70 $\pm$ 0.14 & 0.63 $\pm$ 0.02 & 3.50 $\times$ 10$^{6}$ & 1.20 - 6.50 & 0.24 & 54.56 $\pm$ 18.47 & 20.04 $\pm$ 0.67 \\
43 & Kepler-117 c & 1.84 $\pm$ 0.18 & 1.10 $\pm$ 0.04 & 5.78 $\times$ 10$^{7}$ & 3.90 - 6.70 & -0.04 & 17.21 $\pm$ 14.98 & 2.25 $\pm$ 1.93 \\
44 & Kepler-1514 b & 5.28 $\pm$ 0.22 & 1.11 $\pm$ 0.02 & 5.13 $\times$ 10$^{6}$ & 1.60 - 4.50 & 0.12 & 25.21 $\pm$ 17.86 & 0.82 $\pm$ 0.58 \\
45 & Kepler-16 b & 0.33 $\pm$ 0.02 & 0.75 $\pm$ 0.00 & 4.09 $\times$ 10$^{5}$ & - & -0.30 & 34.94 $\pm$ 3.04 & 47.00 $\pm$ 2.69 \\
46 & Kepler-167 e & 1.01 $\pm$ 0.16 & 0.91 $\pm$ 0.04 & 1.10 $\times$ 10$^{5}$ & 2.50 - 11.50 & 0.02 & 52.49 $\pm$ 26.55 & 10.95 $\pm$ 4.95 \\
47 & Kepler-1704 b & 4.16 $\pm$ 0.29 & 1.07 $\pm$ 0.04 & 9.40 $\times$ 10$^{5}$ & 6.40 - 8.90 & 0.20 & 25.01 $\pm$ 15.23 & 0.86 $\pm$ 0.52 \\
48 & Kepler-289 c & 0.42 $\pm$ 0.05 & 1.03 $\pm$ 0.02 & 6.7 $\times$ 10$^{6}$ & 0.21 - 1.09 & 0.05 & 7.00 $\pm$ 4.92 & 3.28 $\pm$ 2.23 \\
49 & Kepler-30 c & 2.01 $\pm$ 0.16 & 1.10 $\pm$ 0.04 & 1.12 $\times$ 10$^{7}$ & 1.20 - 2.80 & 0.18 & 26.57 $\pm$ 20.91 & 1.94 $\pm$ 1.51 \\
50 & Kepler-34 b & 0.22 $\pm$ 0.01 & 0.76 $\pm$ 0.01 & 1.70 $\times$ 10$^{6}$ & - & -0.07 & 18.43 $\pm$ 2.79 & 22.10 $\pm$ 3.00 \\
51 & Kepler-35 b & 0.13 $\pm$ 0.02 & 0.73 $\pm$ 0.01 & 5.41 $\times$ 10$^{6}$ & - & -0.34 & 12.63 $\pm$ 2.89 & 48.31 $\pm$ 6.98 \\
52 & Kepler-39 b & 17.90 $\pm$ 1.80 & 1.09 $\pm$ 0.08 & 1.20 $\times$ 10$^{8}$ & 0.20 - 6.70 & -0.29 & 103.35 $\pm$ 65.01 & 2.59 $\pm$ 1.60 \\
53 & Kepler-413 b & 0.21 $\pm$ 0.07 & 0.39 $\pm$ 0.01 & 4.03 $\times$ 10$^{6}$ & - & -0.20 & - & - \\
54 & Kepler-419 b & 2.50 $\pm$ 0.30 & 0.96 $\pm$ 0.12 & 4.62 $\times$ 10$^{7}$ & 1.50 - 4.10 & 0.18 & 85.16 $\pm$ 46.17 & 5.11 $\pm$ 2.68 \\
55 & Kepler-432 b & 0.57 $\pm$ 0.02 & 1.15 $\pm$ 0.04 & 1.42 $\times$ 10$^{8}$ & 3.20 - 5.00 & -0.07 & 2.45$^{+2.53}_{-2.45}$ & 1.08$^{+1.11}_{-1.08}$ \\
56 & Kepler-45 b & 0.51 $\pm$ 0.09 & 0.96 $\pm$ 0.11 & 8.82 $\times$ 10$^{7}$ & - & 0.28 & 29.40 $\pm$ 22.94 & 6.74 $\pm$ 4.99 \\
57 & Kepler-539 b & 0.97 $\pm$ 0.29 & 0.75 $\pm$ 0.02 & 5.8 $\times$ 10$^{6}$ & 1.37 - 3.20 & -0.01 & 111.69 $\pm$ 41.30 & 34.16 $\pm$ 2.33 \\
58 & Kepler-553 c & 6.70 $\pm$ 0.43 & 1.03 $\pm$ 0.03 & 8.94 $\times$ 10$^{5}$ & 4.80 - 12.10 & 0.15 & 39.40 $\pm$ 24.33 & 0.94 $\pm$ 0.57 \\
59 & Kepler-75 b & 9.90 $\pm$ 0.50 & 1.03 $\pm$ 0.06 & 1.20 $\times$ 10$^{8}$ & 3.00 - 9.00 & -0.07 & 61.48 $\pm$ 36.54 & 1.64 $\pm$ 0.97 \\
60 & Kepler-849 b & 0.94 $\pm$ 0.20 & 0.72 $\pm$ 0.03 & 3.92 $\times$ 10$^{6}$ & 3.10 - 6.50 & 0.14 & 124.50 $\pm$ 32.64 & 23.81 $\pm$ 1.53 \\
61 & Kepler-87 b & 1.02 $\pm$ 0.03 & 1.20 $\pm$ 0.05 & 1.73 $\times$ 10$^{7}$ & 7.00 - 8.00 & -0.17 & 8.98 $\pm$ 6.12 & 2.92 $\pm$ 1.97 \\
62 & Kepler-9 b & 0.14 $\pm$ 0.01 & 0.74 $\pm$ 0.04 & 6.14 $\times$ 10$^{7}$ & 0.70 - 4.00 & 0.05 & 18.21 $\pm$ 3.40 & 26.72 $\pm$ 4.86 \\
63 & NGTS-11 b & 0.34 $\pm$ 0.08 & 0.82 $\pm$ 0.03 & 1.36 $\times$ 10$^{7}$ & 2.30 - 5.50 & 0.22 & 30.57 $\pm$ 11.95 & 11.76 $\pm$ 2.78 \\
64 & NGTS-20 b & 2.98 $\pm$ 0.15 & 1.07 $\pm$ 0.04 & 5.7 $\times$ 10$^{7}$ & 1.40 - 6.80 & 0.15 & 57.54 $\pm$ 40.06 & 3.07 $\pm$ 2.13 \\
65 & NGTS-30 b & 0.96 $\pm$ 0.06 & 0.93 $\pm$ 0.03 & 5.37 $\times$ 10$^{6}$ & 0.70 - 1.50 & -0.03 & 72.45 $\pm$ 17.97 & 18.15 $\pm$ 4.29 \\
66 & PH2 b & 0.27 $\pm$ 0.07 & 0.83 $\pm$ 0.02 & 1.71 $\times$ 10$^{6}$ & 1.30 - 8.40 & 0.02 & 14.52 $\pm$ 7.80 & 10.94 $\pm$ 4.37 \\
67 & TIC 139270665 b & 0.46 $\pm$ 0.05 & 0.65 $\pm$ 0.02 & 5.60 $\times$ 10$^{7}$ & 1.00 - 7.20 & 0.16 & 93.05 $\pm$ 11.94 & 31.49 $\pm$ 2.03 \\
68 & TIC 172900988 Aa b & 2.63 $\pm$ 0.01 & 1.01 $\pm$ 0.04 & 4.01 $\times$ 10$^{6}$ & 3.00 - 3.20 & 0.34 & 103.01 $\pm$ 46.55 & 4.02 $\pm$ 1.82 \\
69 & TIC 172900988 b & 2.96 $\pm$ 0.02 & 1.00 $\pm$ 0.04 & 6.01 $\times$ 10$^{6}$ & 3.00 - 3.20 & 0.34 & 111.61 $\pm$ 43.91 & 3.87 $\pm$ 1.52 \\
70 & TIC 237913194 b & 1.94 $\pm$ 0.09 & 1.12 $\pm$ 0.05 & 1.11 $\times$ 10$^{8}$ & 4.00 - 7.40 & 0.14 & 25.67 $\pm$ 22.46 & 2.14 $\pm$ 1.87 \\
71 & TIC 279401253 b & 6.14 $\pm$ 0.41 & 1.00 $\pm$ 0.04 & 1.27 $\times$ 10$^{7}$ & 0.40 - 2.20 & 0.20 & 46.33 $\pm$ 22.79 & 1.06 $\pm$ 0.51 \\
72 & TIC 393818343 b & 4.34 $\pm$ 0.15 & 1.09 $\pm$ 0.02 & 9.60 $\times$ 10$^{7}$ & 1.70 - 6.40 & 0.32 & 40.83 $\pm$ 30.34 & 1.02 $\pm$ 0.76 \\
73 & TIC-6432937 b & 3.20 $\pm$ 0.11 & 1.19 $\pm$ 0.03 & 1.32 $\times$ 10$^{8}$ & 2.30 - 12.50 & 0.32 & 12.38 $\pm$ 11.96 & 0.41 $\pm$ 0.40 \\
74 & TOI-1130 c & 1.02 $\pm$ 0.02 & 1.19 $\pm$ 0.13 & 4.03 $\times$ 10$^{7}$ & 3.20 - 5.00 & 0.30 & 35.75 $\pm$ 30.95 & 3.93 $\pm$ 3.40 \\
75 & TOI-1259 A b & 0.44 $\pm$ 0.05 & 1.02 $\pm$ 0.03 & 1.94 $\times$ 10$^{8}$ & 4.00 - 5.50 & -0.50 & 3.64 $\pm$ 3.25 & 5.63 $\pm$ 4.90 \\
76 & TOI-1268 b & 0.30 $\pm$ 0.03 & 0.81 $\pm$ 0.05 & 1.62 $\times$ 10$^{8}$ & 0.11 - 0.38 & 0.36 & 44.14 $\pm$ 8.73 & 14.27 $\pm$ 2.51 \\
77 & TOI-1386 b & 0.15 $\pm$ 0.02 & 0.54 $\pm$ 0.02 & 4.70 $\times$ 10$^{7}$ & 1.10 - 6.80 & 0.16 & 33.11 $\pm$ 4.56 & 34.41 $\pm$ 1.10 \\
78 & TOI-1478 b & 0.85 $\pm$ 0.05 & 1.06 $\pm$ 0.04 & 1.63 $\times$ 10$^{8}$ & 5.20 - 12.20 & 0.08 & 9.08 $\pm$ 7.88 & 1.98 $\pm$ 1.71 \\
79 & TOI-1670 c & 0.63 $\pm$ 0.08 & 0.99 $\pm$ 0.03 & 4.99 $\times$ 10$^{7}$ & 2.10 - 2.96 & 0.09 & 14.92 $\pm$ 8.76 & 4.23 $\pm$ 2.29 \\
80 & TOI-181 b & 0.15 $\pm$ 0.02 & 0.63 $\pm$ 0.01 & 1.44 $\times$ 10$^{8}$ & 0.50 - 10.30 & 0.27 & 26.31 $\pm$ 3.52 & 21.75 $\pm$ 1.32 \\
81 & TOI-1811 b & 0.97 $\pm$ 0.08 & 0.99 $\pm$ 0.02 & 1.94 $\times$ 10$^{8}$ & 1.90 - 10.80 & 0.31 & 21.34 $\pm$ 13.02 & 2.42 $\pm$ 1.44 \\
82 & TOI-1899 b & 0.67 $\pm$ 0.04 & 0.99 $\pm$ 0.03 & 4.29 $\times$ 10$^{6}$ & 2.30 - 11.60 & 0.28 & 9.39 $\pm$ 7.43 & 1.63 $\pm$ 1.28 \\
83 & TOI-199 b & 0.17 $\pm$ 0.02 & 0.81 $\pm$ 0.01 & 3.45 $\times$ 10$^{6}$ & 0.20 - 2.00 & 0.22 & 15.23 $\pm$ 3.20 & 12.11 $\pm$ 1.94 \\
84 & TOI-201 b & 0.42 $\pm$ 0.04 & 1.01 $\pm$ 0.01 & 3.88 $\times$ 10$^{7}$ & 0.38 - 1.33 & 0.24 & 9.59 $\pm$ 4.48 & 2.93 $\pm$ 1.30 \\
85 & TOI-2010 b & 1.29 $\pm$ 0.06 & 1.05 $\pm$ 0.03 & 5.80 $\times$ 10$^{6}$ & 0.60 - 4.10 & 0.17 & 20.03 $\pm$ 14.24 & 2.37 $\pm$ 1.68 \\
86 & TOI-2134 c & 0.13 $\pm$ 0.02 & 0.65 $\pm$ 0.04 & 1.98 $\times$ 10$^{6}$ & 1.10 - 9.30 & 0.12 & 19.35 $\pm$ 5.21 & 24.66 $\pm$ 4.36 \\
87 & TOI-2180 b & 2.75 $\pm$ 0.08 & 1.01 $\pm$ 0.02 & 5.6 $\times$ 10$^{6}$ & 6.80 - 9.60 & 0.25 & 73.88 $\pm$ 28.93 & 3.36 $\pm$ 1.31 \\
88 & TOI-2338 b & 5.98 $\pm$ 0.21 & 1.00 $\pm$ 0.02 & 5.25 $\times$ 10$^{7}$ & 5.00 - 9.00 & 0.22 & 51.37 $\pm$ 19.36 & 1.16 $\pm$ 0.43 \\
89 & TOI-2373 b & 9.30 $\pm$ 0.20 & 0.93 $\pm$ 0.02 & 1.21 $\times$ 10$^{8}$ & 4.20 - 7.60 & 0.30 & 90.62 $\pm$ 28.43 & 1.10 $\pm$ 0.33 \\
90 & TOI-2447 b & 0.39 $\pm$ 0.03 & 0.86 $\pm$ 0.01 & 1.12 $\times$ 10$^{7}$ & 1.10 - 3.10 & 0.18 & 31.58 $\pm$ 4.76 & 11.91 $\pm$ 1.36 \\
91 & TOI-2525 c & 0.66 $\pm$ 0.03 & 0.90 $\pm$ 0.01 & 8.32 $\times$ 10$^{6}$ & 1.39 - 8.29 & 0.14 & 36.34 $\pm$ 7.39 & 8.99 $\pm$ 1.71 \\
92 & TOI-2589 b & 3.50 $\pm$ 0.10 & 1.08 $\pm$ 0.03 & 1.51 $\times$ 10$^{7}$ & 9.00 - 13.00 & 0.12 & 31.77 $\pm$ 26.19 & 1.55 $\pm$ 1.27 \\
93 & TOI-3235 b & 0.67 $\pm$ 0.03 & 1.02 $\pm$ 0.04 & 3.02 $\times$ 10$^{7}$ & - & 0.26 & 11.84 $\pm$ 9.59 & 2.17 $\pm$ 1.75 \\
94 & TOI-3629 b & 0.24 $\pm$ 0.02 & 0.74 $\pm$ 0.01 & 5.79 $\times$ 10$^{7}$ & 5.00 - 16.40 & 0.55 & 28.08 $\pm$ 3.77 & 7.32 $\pm$ 0.65 \\
95 & TOI-3693 b & 1.02 $\pm$ 0.23 & 1.12 $\pm$ 0.03 & 9.28 $\times$ 10$^{7}$ & 0.90 - 6.80 & 0.07 & 6.60 $\pm$ 6.50 & 1.04 $\pm$ 0.97 \\
96 & TOI-3714 b & 0.69 $\pm$ 0.03 & 0.99 $\pm$ 0.02 & 7.88 $\times$ 10$^{7}$ & 6.00 - 17.40 & 0.39 & 5.89 $\pm$ 4.69 & 0.78 $\pm$ 0.61 \\
97 & TOI-3757 b & 0.27 $\pm$ 0.03 & 1.07 $\pm$ 0.04 & 7.48 $\times$ 10$^{7}$ & 2.60 - 11.60 & 0.00 & 2.00 $\pm$ 1.94 & 1.65 $\pm$ 1.57 \\
98 & TOI-3984 A b & 0.14 $\pm$ 0.03 & 0.71 $\pm$ 0.02 & 2.33 $\times$ 10$^{7}$ & 0.70 - 5.10 & 0.18 & 19.87 $\pm$ 4.41 & 21.12 $\pm$ 2.34 \\
99 & TOI-4010 d & 0.12 $\pm$ 0.01 & 0.55 $\pm$ 0.01 & 4.02 $\times$ 10$^{7}$ & 3.00 - 9.20 & 0.37 & 25.38 $\pm$ 2.47 & 20.24 $\pm$ 0.72 \\
100 & TOI-4127 b & 2.30 $\pm$ 0.11 & 1.10 $\pm$ 0.04 & 2.95 $\times$ 10$^{7}$ & 2.70 - 6.90 & 0.14 & 25.25 $\pm$ 20.87 & 1.78 $\pm$ 1.46 \\
101 & TOI-4201 b & 2.48 $\pm$ 0.09 & 1.22 $\pm$ 0.04 & 6.66 $\times$ 10$^{7}$ & 0.70 - 2.00 & 0.52 & 13.40 $\pm$ 12.63 & 0.37 $\pm$ 0.34 \\
102 & TOI-4406 b & 0.30 $\pm$ 0.03 & 1.00 $\pm$ 0.02 & 7.56 $\times$ 10$^{7}$ & 2.20 - 3.60 & 0.10 & 2.15 $\pm$ 1.80 & 1.25 $\pm$ 1.02 \\
103 & TOI-4515 b & 2.00 $\pm$ 0.05 & 1.09 $\pm$ 0.04 & 5.68 $\times$ 10$^{7}$ & 1.00 - 1.40 & 0.05 & 39.14 $\pm$ 27.16 & 3.91 $\pm$ 2.71 \\
104 & TOI-4562 b & 2.30 $\pm$ 0.47 & 1.12 $\pm$ 0.01 & 3.80 $\times$ 10$^{6}$ & - & 0.08 & 7.06 $\pm$ 6.66 & 0.48 $\pm$ 0.43 \\
105 & TOI-4582 b & 0.53 $\pm$ 0.05 & 0.94 $\pm$ 0.11 & 1.22 $\times$ 10$^{8}$ & - & 0.17 & 32.55 $\pm$ 23.36 & 9.29 $\pm$ 6.55 \\
106 & TOI-4860 b & 0.27 $\pm$ 0.01 & 0.78 $\pm$ 0.04 & 5.47 $\times$ 10$^{7}$ & - & 0.27 & 28.21 $\pm$ 6.38 & 12.48 $\pm$ 2.80 \\
107 & TOI-5153 b & 3.26 $\pm$ 0.17 & 1.06 $\pm$ 0.04 & 1.52 $\times$ 10$^{8}$ & 4.40 - 6.40 & 0.12 & 64.69 $\pm$ 44.20 & 3.38 $\pm$ 2.30 \\
108 & TOI-5176 A b & 0.62 $\pm$ 0.05 & 0.94 $\pm$ 0.03 & 4.68 $\times$ 10$^{6}$ & 2.60 - 11.80 & 0.00 & 16.96 $\pm$ 11.05 & 6.12 $\pm$ 3.91 \\
109 & TOI-519 b & 0.46 $\pm$ 0.09 & 1.03 $\pm$ 0.03 & 7.14 $\times$ 10$^{7}$ & 0.80 - 10.00 & 0.27 & 5.92 $\pm$ 5.22 & 1.44 $\pm$ 1.21 \\
110 & TOI-5205 b & 1.08 $\pm$ 0.06 & 1.04 $\pm$ 0.03 & 6.56 $\times$ 10$^{7}$ & - & 0.00 & 13.55 $\pm$ 10.46 & 2.81 $\pm$ 2.15 \\
111 & TOI-5293 A b & 0.54 $\pm$ 0.07 & 1.06 $\pm$ 0.04 & 4.74 $\times$ 10$^{7}$ & 0.70 - 5.10 & -0.03 & 7.21 $\pm$ 6.27 & 3.11 $\pm$ 2.63 \\
112 & TOI-530 b & 0.40 $\pm$ 0.10 & 0.83 $\pm$ 0.06 & 2.37 $\times$ 10$^{7}$ & - & 0.38 & 33.45 $\pm$ 18.33 & 7.71 $\pm$ 3.47 \\
113 & TOI-532 b & 0.19 $\pm$ 0.03 & 0.52 $\pm$ 0.02 & 1.23 $\times$ 10$^{8}$ & 2.30 - 11.50 & 0.34 & 45.02 $\pm$ 7.78 & 23.95 $\pm$ 0.75 \\
114 & TOI-5344 b & 0.41 $\pm$ 0.04 & 0.95 $\pm$ 0.02 & 5.15 $\times$ 10$^{7}$ & 5.10 - 16.70 & 0.42 & 5.41 $\pm$ 4.04 & 1.08 $\pm$ 0.78 \\
115 & TOI-5398 b & 0.18 $\pm$ 0.02 & 0.92 $\pm$ 0.04 & 1.79 $\times$ 10$^{8}$ & 0.50 - 0.80 & 0.09 & 14.02 $\pm$ 3.29 & 13.90 $\pm$ 3.08 \\
116 & TOI-5542 b & 1.32 $\pm$ 0.10 & 1.01 $\pm$ 0.04 & 1.29 $\times$ 10$^{7}$ & 7.20 - 12.90 & -0.21 & 25.27 $\pm$ 18.33 & 6.91 $\pm$ 4.94 \\
117 & TOI-6034 b & 0.80 $\pm$ 0.08 & 1.06 $\pm$ 0.04 & 5.88 $\times$ 10$^{7}$ & 2.60 - 12.00 & 0.00 & 9.75 $\pm$ 8.56 & 2.67 $\pm$ 2.31 \\
118 & TOI-6383 b & 0.89 $\pm$ 0.09 & 0.90 $\pm$ 0.04 & 6.79 $\times$ 10$^{7}$ & - & 0.20 & 62.07 $\pm$ 22.34 & 9.86 $\pm$ 3.26 \\
119 & TOI-762 A b & 0.25 $\pm$ 0.04 & 0.74 $\pm$ 0.02 & 2.16 $\times$ 10$^{7}$ & 3.80 - 14.80 & 0.36 & 27.94 $\pm$ 6.82 & 10.90 $\pm$ 1.35 \\
120 & WASP-105 b & 1.80 $\pm$ 0.10 & 0.96 $\pm$ 0.03 & 1.18 $\times$ 10$^{8}$ & - & 0.28 & 96.09 $\pm$ 31.74 & 6.29 $\pm$ 2.03 \\
121 & WASP-107 b & 0.12 $\pm$ 0.01 & 0.94 $\pm$ 0.02 & 6.80 $\times$ 10$^{7}$ & 2.70 - 4.10 & 0.02 & 2.77 $\pm$ 1.21 & 4.98 $\pm$ 2.20 \\
122 & WASP-11 b & 0.79 $\pm$ 0.17 & 1.11 $\pm$ 0.10 & 1.84 $\times$ 10$^{8}$ & 4.60 - 13.40 & 0.00 & 24.72 $\pm$ 22.66 & 6.64 $\pm$ 5.73 \\
123 & WASP-130 b & 1.23 $\pm$ 0.04 & 0.89 $\pm$ 0.03 & 1.10 $\times$ 10$^{8}$ & - & 0.26 & 103.66 $\pm$ 22.78 & 10.40 $\pm$ 2.25 \\
124 & WASP-132 b & 0.41 $\pm$ 0.03 & 0.87 $\pm$ 0.03 & 7.78 $\times$ 10$^{7}$ & - & 0.22 & 24.88 $\pm$ 8.59 & 8.19 $\pm$ 2.70 \\
125 & WASP-139 b & 0.12 $\pm$ 0.02 & 0.80 $\pm$ 0.05 & 1.62 $\times$ 10$^{8}$ & - & 0.20 & 10.50 $\pm$ 3.53 & 12.59 $\pm$ 3.95 \\
126 & WASP-148 b & 0.29 $\pm$ 0.02 & 0.76 $\pm$ 0.02 & 1.43 $\times$ 10$^{8}$ & 0.90 - 7.60 & 0.10 & 36.46 $\pm$ 4.66 & 22.71 $\pm$ 2.30 \\
127 & WASP-162 b & 5.20 $\pm$ 0.20 & 1.00 $\pm$ 0.05 & 1.57 $\times$ 10$^{8}$ & 10.62 - 15.32 & 0.28 & 35.65 $\pm$ 19.06 & 0.81 $\pm$ 0.43 \\
128 & WASP-59 b & 0.86 $\pm$ 0.05 & 0.78 $\pm$ 0.07 & 4.38 $\times$ 10$^{7}$ & 0.10 - 1.20 & -0.15 & 114.17 $\pm$ 19.09 & 42.04 $\pm$ 6.63 \\
129 & WASP-69 b & 0.26 $\pm$ 0.02 & 1.06 $\pm$ 0.05 & 1.94 $\times$ 10$^{8}$ & - & 0.14 & 2.78 $\pm$ 2.54 & 1.72 $\pm$ 1.56 \\
130 & WASP-8 b & 2.13 $\pm$ 0.08 & 1.04 $\pm$ 0.03 & 1.66 $\times$ 10$^{8}$ & 3.00 - 5.00 & 0.17 & 48.02 $\pm$ 26.12 & 3.42 $\pm$ 1.85 \\
131 & WASP-80 b & 0.54 $\pm$ 0.04 & 1.00 $\pm$ 0.03 & 1.6 $\times$ 10$^{8}$ & - & -0.13 & 7.69 $\pm$ 6.04 & 4.29 $\pm$ 3.33 \\
132 & WASP-84 b & 0.69 $\pm$ 0.03 & 0.98 $\pm$ 0.03 & 9.41 $\times$ 10$^{7}$ & 0.50 - 3.70 & 0.09 & 25.56 $\pm$ 11.07 & 6.78 $\pm$ 2.90 \\
133 & Wendelstein-1 b & 0.59 $\pm$ 0.15 & 1.03 $\pm$ 0.01 & 1.40 $\times$ 10$^{8}$ & - & - & 3.64 $\pm$ 3.22 & - \\
\end{longtable}
\end{center}
\twocolumn

\appendix

\section{Exoplanetary sample}
\label{sec:appendix_exoplanetary_sample}

\begin{figure}[h]
    \includegraphics[width=\columnwidth]{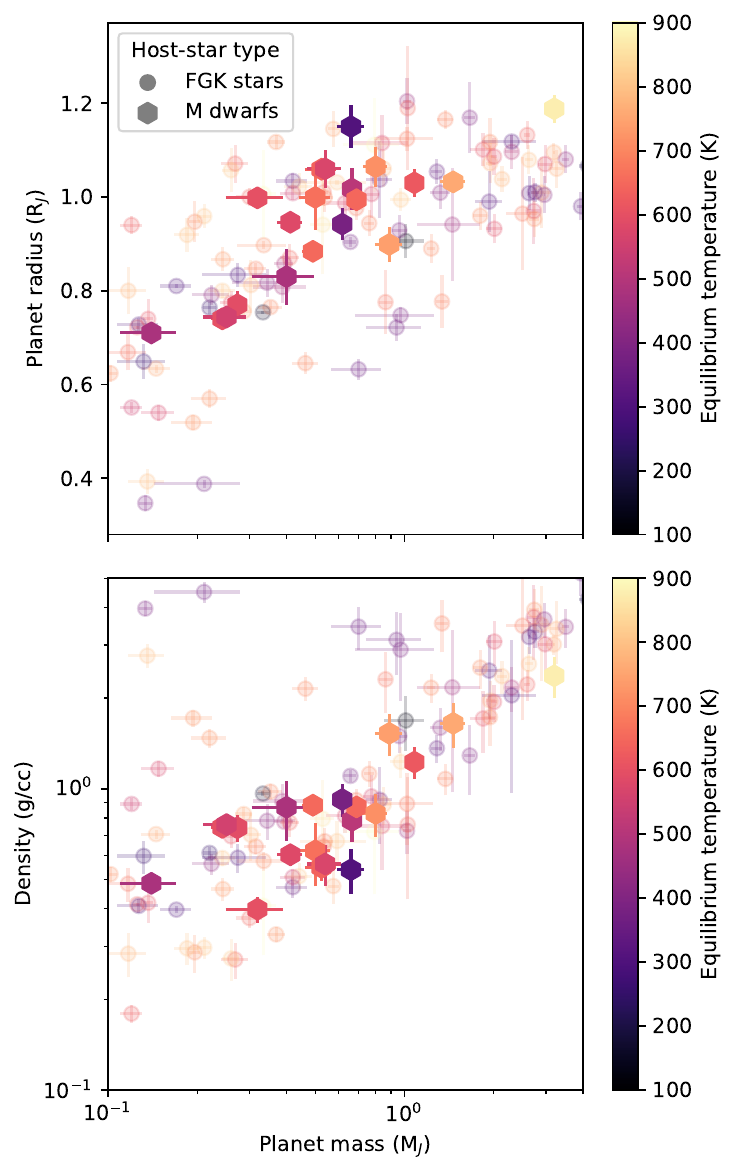}
    \caption{Mass-radius (top) and mass-density relations (bottom) for giant planets around FGK stars (circles) and M-dwarfs (hexagons). The colours indicate the equilibrium temperature.}
    \label{fig:mass_radius_density}
\end{figure}

Figure \ref{fig:mass_radius_density} shows the warm giant planet sample in the mass-radius and mass-density space. The full dataset included 110 and 20 planets around FGK and M-dwarf stars, respectively. There is no immediately obvious difference between the two populations evident. This highlights the need for thermal evolution models to characterise warm giant planets.

In the sample, there were 11 low-mass planets around FGK stars (< 0.9 M$_J$) with very high densities (> 1.1 g/cm$^3$). Here, we briefly discuss whether they could explain the differences in the mass-metallicity trend that we observed. Four planets were too dense for a successful interior characterisation with our models and were not included in the analysis. To explain the observations, these planets would need to consist of large fractions of rocks or iron. For the other seven planets, three were excluded from our preferred sample due to being less massive than 0.3 M$_J$. The remaining four planets were used in the fitting procedure. Two were found to be quite metal-rich ($Z / Z_* \simeq 30 - 40$), and two were moderately metal-rich ($Z / Z_* \simeq 10 - 15$). Given the large number of giant planets around FGK stars in our dataset, we don't expect the two metal-rich planets with anomalously high densities to significantly influence the inferred $M$-$M_z$ and $M$-$Z / Z_*$ relations.

\begin{figure}[h]
    \includegraphics[width=\columnwidth]{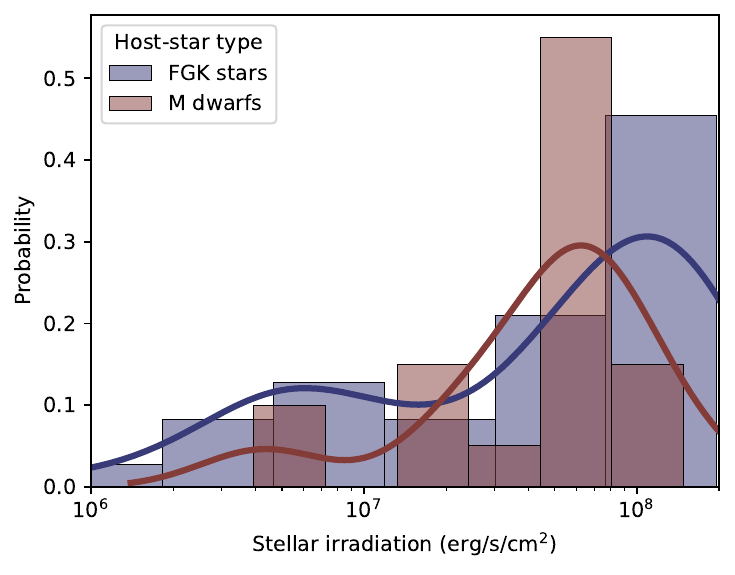}
    \caption{Histograms and kernel density estimates of the stellar irradiation for the giant planets around FGK stars (blue) and M dwarfs (red).}
    \label{fig:stellar_irradiation}
\end{figure}

Figure \ref{fig:stellar_irradiation} shows the distribution of the stellar irradiation for the two populations. The distributions are similar, but the peak for the planets around FGK stars is around $I_* = 10^8$ compared to around $5 \times 10^7$ erg/s/cm$^2$ for M dwarfs. Therefore, the former have higher equilibrium temperatures than the latter. Since M-dwarf stars are colder, this also means that the planets around them have very short orbital periods.

\section{$M_p$-$Z / Z_*$ and $M_p$-$M_z$ relations for the full sample}
\label{sec:appendix_full_sample}

In the full sample ($0.1 \leq M_p (M_J) \leq 10$), there were 103 successful bulk-metallicity retrievals for FGK stars and 20 for M-dwarf stars. The reasons for an unsuccessful retrieval are that the planet's density is too high or low at its suggested age. The inferred mass-metallicity relations were

\begin{equation}
    \label{eq:mass_metallicity_full}
    Z / Z_* = 
    \begin{cases}
        (6.35 \pm 0.39) M_p^{(-0.71 \pm 0.04)} & \text{FGK stars} \, ,\\
        (1.22 \pm 0.30) M_p^{(-1.42 \pm 0.19)} & \text{M dwarfs} \, ,
    \end{cases}
\end{equation}

\noindent
For the heavy-element masses, the relations were

\begin{equation}
    \label{eq:mass_mass_full}
    M_z = 
    \begin{cases}
        (42.64 \pm 2.84) M_p^{(0.39 \pm 0.03)} & \text{FGK stars} \, ,\\
        (16.93 \pm 3.80) M_p^{(-0.29 \pm 0.18)} & \text{M dwarfs} \, ,
    \end{cases}
\end{equation}

\begin{figure}[h]
    \includegraphics[width=\columnwidth]{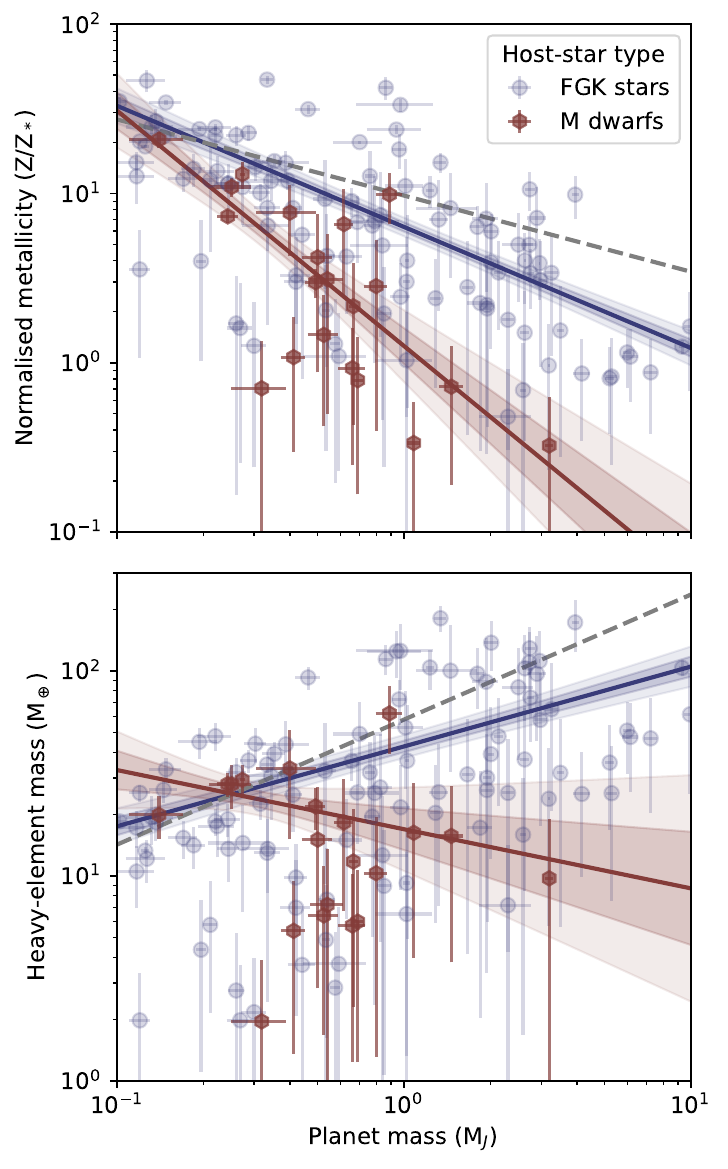}
    \caption{Same as Figure \ref{fig:metallicity_limited}, but the fits were for the full planet sample with $0.1 \leq M_p (M_J) \leq 10$.}
    \label{fig:metallicity_full}
\end{figure}

\noindent
Figure \ref{fig:metallicity_full} shows the inferred normalised metallicities, heavy-element masses, and the two fits. As before, the fit parameters are tabulated and their posterior distributions are shown in Appendix \ref{sec:appendix_fit_parameters}.\noindent
where $M_z$ is in Earth units. 

Compared to the results in Section \ref{sec:results_mass_limited_sample}, the larger mass range yielded significantly different $M_p$-$Z / Z_*$ and $M_p$-$M_z$ relations. We find a qualitatively similar inverse trend between the masses and bulk metallicities. However, for both populations, the slopes are significantly steeper and do not agree with the results from \citet{2016ApJ...831...64T}. \editone{As already discussed in Section \ref{sec:results_mass_limited_sample}, we generally infer a lower metallicity than Thorngren et al., mainly due to the more realistic hydrogen-helium equation of state used in our models.} The $M_p$-$Z / Z_*$ relation also declines more rapidly for the giant planets around M dwarfs and predicts a lower enrichment for all masses for these planets. The main conclusion is consistent with our previous analysis: i.e., giant planets around M dwarfs are less metal-rich. This is particularly true for the planets more massive than about a Saturn mass. 

For the $M_p$-$M_z$ relations, there are larger differences compared to the mass-limited sample and the trends show different behaviours for the two populations: The heavy-element mass increases with mass for the giant planets around FGK stars, and decreases for those around M dwarfs. As for the mass-metallicity trend, our inferred slopes disagree with the results from \citet{2016ApJ...831...64T}. However, the uncertainty on the slope in the latter case is large, and a flat or positive slope is consistent within two $\sigma$.

Looking closer at the inferred composition reveals a potential issue with the fit for the planets around M dwarfs. The low-mass planets are metal-rich, while those above about a Saturn mass are generally metal-poor. Therefore, to fit the low- and high-mass planets the slope becomes very steep, resulting in a negative $M_p$-$M_z$ trend. As discussed previously (see Section \ref{sec:results_mass_limited_sample}), giant planes may better be classified as having masses beyond about $0.3 M_J$. We therefore suggest that the results presented in Section \ref{sec:results_mass_limited_sample} for the mass-limited sample are more realistic and should be preferred.

\section{Fit parameters and posterior distributions}
\label{sec:appendix_fit_parameters}

\begin{figure}[h]
    \includegraphics[width=\columnwidth]{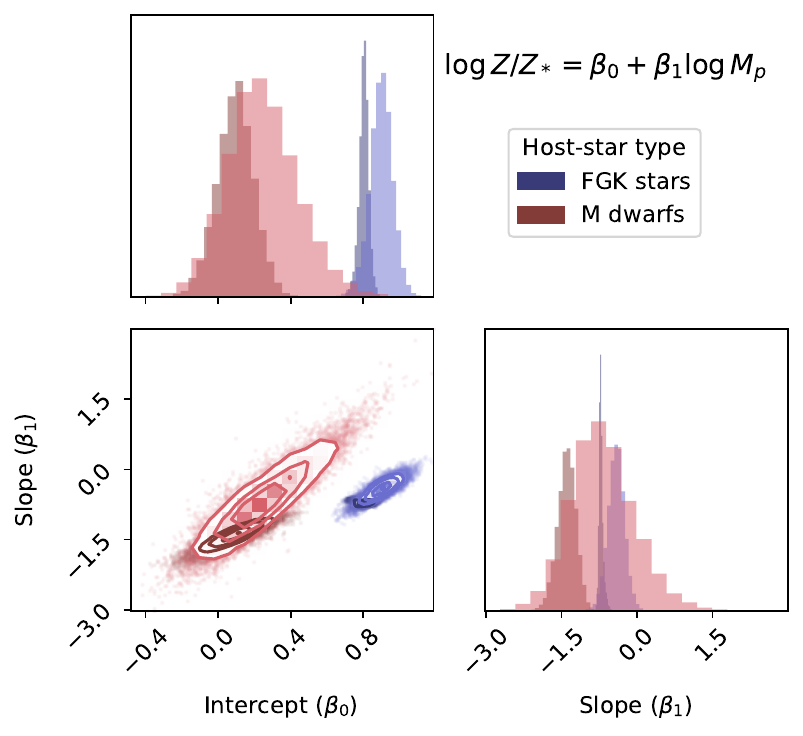}
    \caption{Posterior distributions of the inferred intercept ($\beta_0)$ and slope ($\beta_1)$ for the $M_p$-$Z / Z_*$ relation. The results are coloured depending on the host-star type (blue for FGK stars and red for M dwarfs). The lighter-shaded results are for the fit for the mass-limited sample $0.3 \leq M_p (M_J) \leq 2$.}
    \label{fig:corner_metallicity}
\end{figure}

\begin{figure}[h]
    \includegraphics[width=\columnwidth]{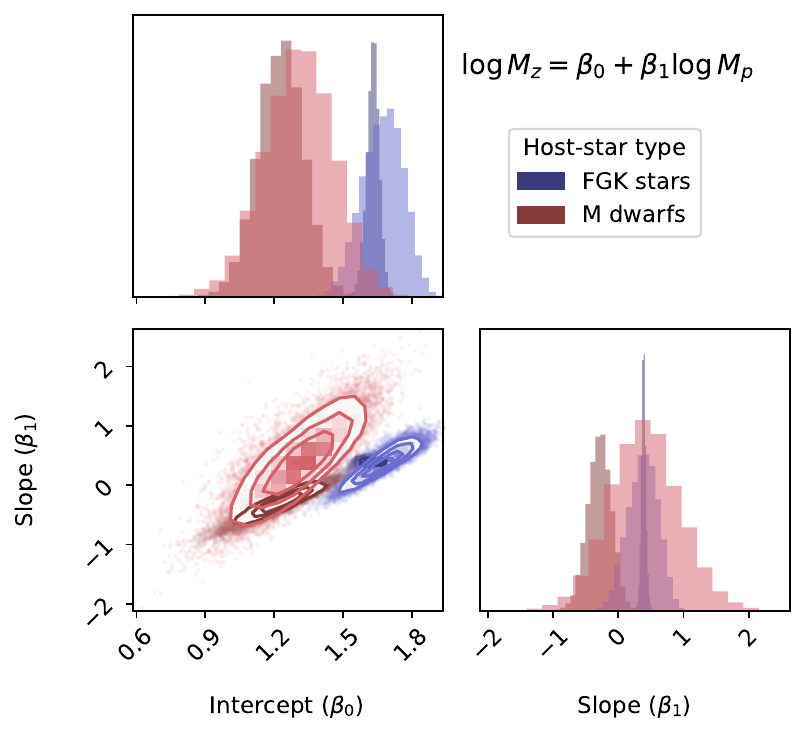}
    \caption{Same as Figure \ref{fig:corner_metallicity}, but for the $M_p$-$M_z$ relation.}
    \label{fig:corner_mass}
\end{figure}

Tables \ref{tab:mass_metallicity} and \ref{tab:mass_mass}  list the parameters for the $M_p$-$Z / Z_*$ and $M_p$-$M_z$ relations, and Figures \ref{fig:corner_metallicity} and \ref{fig:corner_mass} show the posterior distributions of the inferred intercept ($\beta_0$) and slopes ($\beta_1$). The fit parameters are defined in Equations \ref{eq:linear} and \ref{eq:power_law} and are listed for the two different mass ranges.

\begin{table}
    \centering
    \caption{Fit parameters for the mass-metallicity relations. The planetary mass for the fitting ranged between 0.1 and 10 and 0.3 to 2 $M_J$.}
    \begin{tabular}{cccccc}
        Stars & $M_p$ & $\beta_0$ & $\beta_1$ & $\beta_2$ \\
        \hline
        FGK & 0.3 - 2 & $0.90 \pm 0.06$ & $-0.41 \pm 0.17$ & $8.01 \pm 1.19$ \\
        M & 0.3 - 2 & $0.23 \pm 0.19$ &$-0.74 \pm 0.60$ & $1.71 \pm 0.73$ \\
        FGK & 0.1 - 10 & $0.80 \pm 0.0$3 & $-0.71 \pm 0.04$ & $6.35 \pm 0.39$ \\
        M & 0.1 - 10 & $0.08 \pm 0.11$ & $-1.42 \pm 0.19$ & $1.22 \pm 0.30$ \\
    \end{tabular}
    \label{tab:mass_metallicity}
\end{table}

\begin{table}
    \centering
    \caption{Same as Table \ref{tab:mass_metallicity}, but for the relations between the planetary mass $M_p (M_J)$ and the heavy-element mass $M_z (M_\oplus)$}
    \begin{tabular}{ccccc}
        Stars & $M_p$ & $\beta_0$ & $\beta_1$ & $\beta_2$ \\
        \hline
        FGK & 0.3 - 2 & $1.67 \pm 0.09$ & $0.37 \pm 0.24$ & $46.81 \pm 9.86$ \\
        M & 0.3 - 2 & $1.30 \pm 0.15$ & $0.40 \pm 0.53$ & $20.17 \pm 6.83$ \\
        FGK & 0.1 - 10 & $1.63 \pm 0.03$ & $0.39 \pm 0.03$ & $42.64 \pm 2.84$ \\
        M & 0.1 - 10 & $1.23 \pm 0.10$ & $-0.29 \pm 0.18$ & $16.93 \pm 3.80$ \\
    \end{tabular}
    \label{tab:mass_mass}
\end{table}

These results demonstrate that for both the relations the $\beta_0$-$\beta_1$ posteriors are distinct. They depend on the host-star type as well as the mass range of the sample. For the full mass range ($0.1 \leq M_p (M_J) \leq 10$), the slopes are incompatible for the two populations. For the smaller mass range ($0.3 \leq M_p (M_J) \leq 2$), the slopes for the planets around M-dwarf stars are compatible with those around FGK stars. However, they have a large uncertainty. In all the cases, the intercepts ($\beta_0$) are inconsistent for the two populations within more than two $\sigma$. We therefore concluded that planets around M-dwarfs are metal-poor compared to those around FGK stars.

\section{Influence of the uniform age priors for unknown M dwarf ages}
\label{sec:appendix_age_prior_test}

As stated in Section \ref{sec:methods}, whenever a stellar age estimate was unavailable, we used a uniform prior from 1 to 10 Gyr. This was the case for 8 out of 20 M dwarfs. Here, we investigate how this influences our results. To isolate the effect of the age priors, we focused on the heavy-element mass since its determination does not rely on stellar metallicity measurements. For both the full and mass-limited samples we re-inferred the heavy-element masses of the M-dwarf planets twice more. Once with a young age uniform prior of 1 to 3 Gyr, and once with an old age prior of 7 to 10 Gyr.

\begin{figure}[h]
    \includegraphics[width=\columnwidth]{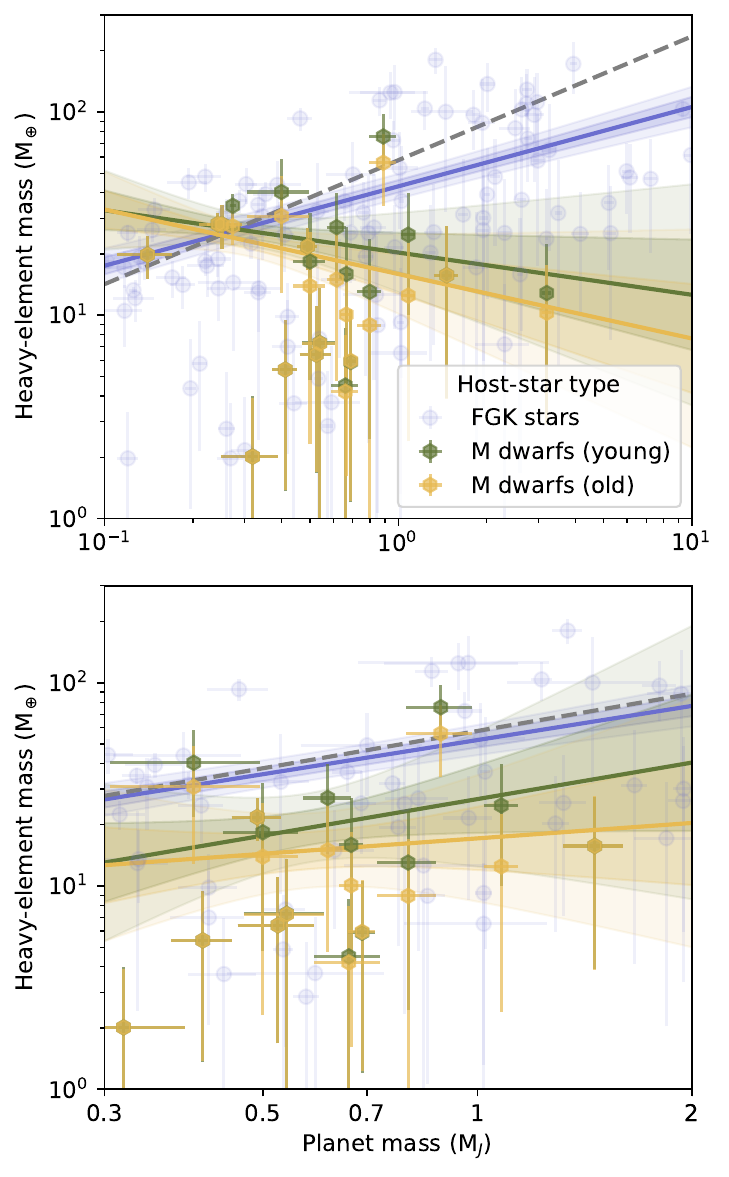}
    \caption{Heavy-element mass as a function of planet mass for the full ($0.1 \leq M_p (M_J) \leq 10$, top) and limited samples ($0.3 \leq M_p (M_J) \leq 2$, bottom). The scatter points and error bars show the inferred heavy-element mass by the thermal evolution models. Solid lines show the fits constructed by Bayesian regression, and the shaded contours are the one and two $\sigma$ uncertainties. Planets around FGK stars are depicted in blue, while planets around M-dwarfs are orange and green depending on the age prior that was used (see text for details).}
    \label{fig:heavy_element_mass_age_prior_test}
\end{figure}

The resulting data and linear fits are shown in Figure \ref{fig:heavy_element_mass_age_prior_test}. Younger age priors lead to higher inferred heavy-element for planets with unknown ages. This shifts the $M$-$M_z$ relation more towards that of the FGK stars. If the eight planets with unknown ages were indeed all very young (1 to 3 Gyr), it would decrease the likelihood that the two planet populations are different concerning their heavy-element masses. Previously, the $M$-$M_z$ relations were inconsistent to within about two $\sigma$, and as Figure \ref{fig:heavy_element_mass_age_prior_test} shows this would be reduced to about one $\sigma$ for the young-age prior. The overall $M_p$-$Z / Z_*$ and $M_p$-$M_z$ trends are also consistent with what was presented in Section \ref{sec:results_mass_limited_sample} and Appendix \ref{sec:appendix_full_sample}. We note, however, that it is very unlikely that all of these planets are this young, and therefore our nominal results are more likely.

\section{Correlations with planetary and stellar properties}
\label{sec:appendix_correlations}

% Residual planetary normalised metallicity
% Significant correlations:
% FGK stars
%   * Stellar mass (full sample): tau = 0.14, p = 4.02e-02
%   * Flux (limited sample): tau = -0.23, p = 1.90e-02
%   * Orbital period (full sample): tau = 0.16, p = 1.70e-02
%   * Semi-major axis (full sample): tau = 0.16, p = 1.57e-02
%   * Eccentricity (full sample): tau = 0.15, p = 2.74e-02
% M-dwarf stars:
%   * None found

% Residual planetary heavy-element mass
% Significant correlations:
% FGK stars
%   * Stellar age (limited sample): tau = -0.21, p = 4.51e-02
%   * Stellar mass (full sample): tau = 0.14, p = 3.03e-02
%   * Flux (limited sample): tau = -0.20, p = 3.73e-02
%   * Orbital period (full sample): tau = 0.15, p = 2.50e-02
%   * Semi-major axis (full sample): tau = 0.15, p = 2.36e-02
%   * Eccentricity (full sample): tau = 0.21, p = 1.97e-03
% M-dwarf stars:
%   * None found

\editone{In this section, we test whether our predictions for the residual normalised metallicity and heavy-element mass are correlated with the stellar age, stellar mass, planetary instellation flux, orbital period, semi-major axis and eccentricity. Similar to our previous analyses, we split the data into planets around FGK and M-dwarf stars and the full and limited mass ranges (see Section \ref{sec:results_planet_selection}). To ensure that we are testing for correlations not accounted for in our fits, we used the residuals, which are the ratios of the calculated values and the ones predicted from the best fits (see \ref{sec:results_correlations}. Similar to the results presented in Section \ref{sec:results_correlations}, we calculated Kendall's tau rank correlation coefficient and its associated p-value for each pairing.}

\editone{For giant planets around M-dwarf stars we did not find any statistically significant correlations. For giant planets around FGK stars with $0.1 \leq M \leq 10 M_J$, however, the residual normalised metallicity was found to be slightly correlated with the stellar mass ($\tau = 0.14$, $p = 0.05$), the orbital period ($\tau = 0.16$, $p = 0.02$), the semi-major axis ($\tau = 0.16$, $p = 0.02$), and the eccentricity ($\tau = 0.15$, $p = 0.03$). For the mass-limited sample, there was also a negative correlation with the instellation flux ($\tau = -0.23$, $p = 0.02$)}.

\editone{For the full mass range, the residual heavy-element mass was similarly correlated with the stellar mass ($\tau = 0.14$, $p = 0.03$), the orbital period ($\tau = 0.15$, $p = 0.03$), the semi-major axis ($\tau = 0.15$, $p = 0.03$) and the eccentricity ($\tau = 0.21$, $p = 0.002$). For this sample, we also obtained a negative correlation with the stellar age ($\tau = -0.21$, $p = 0.04$), and a similar correlation with the instellation flux for the mass-limited sample ($\tau = -0.20$, $p = 0.03$).}

\editone{Overall, these inferred correlations are generally weak with relatively large p-values for the FGK sample. The lack of correlations for the planets around M-dwarf stars is probably caused by the limited data. Future research could re-visit this topic once more data are available.} 
%The most significant correlation we obtained was the one between the residual heavy-element mass and the eccentricity. 
%{\it delete: However, clearly all parameters we investigated are much less related to the metallicity and heavy-element mass than the planetary mass. 
%We also note that while we did not find any correlations for the planets around M-dwarf stars, this may also be due to the lack of data.}

\begin{figure}
    \centering
    \includegraphics[width=\columnwidth]{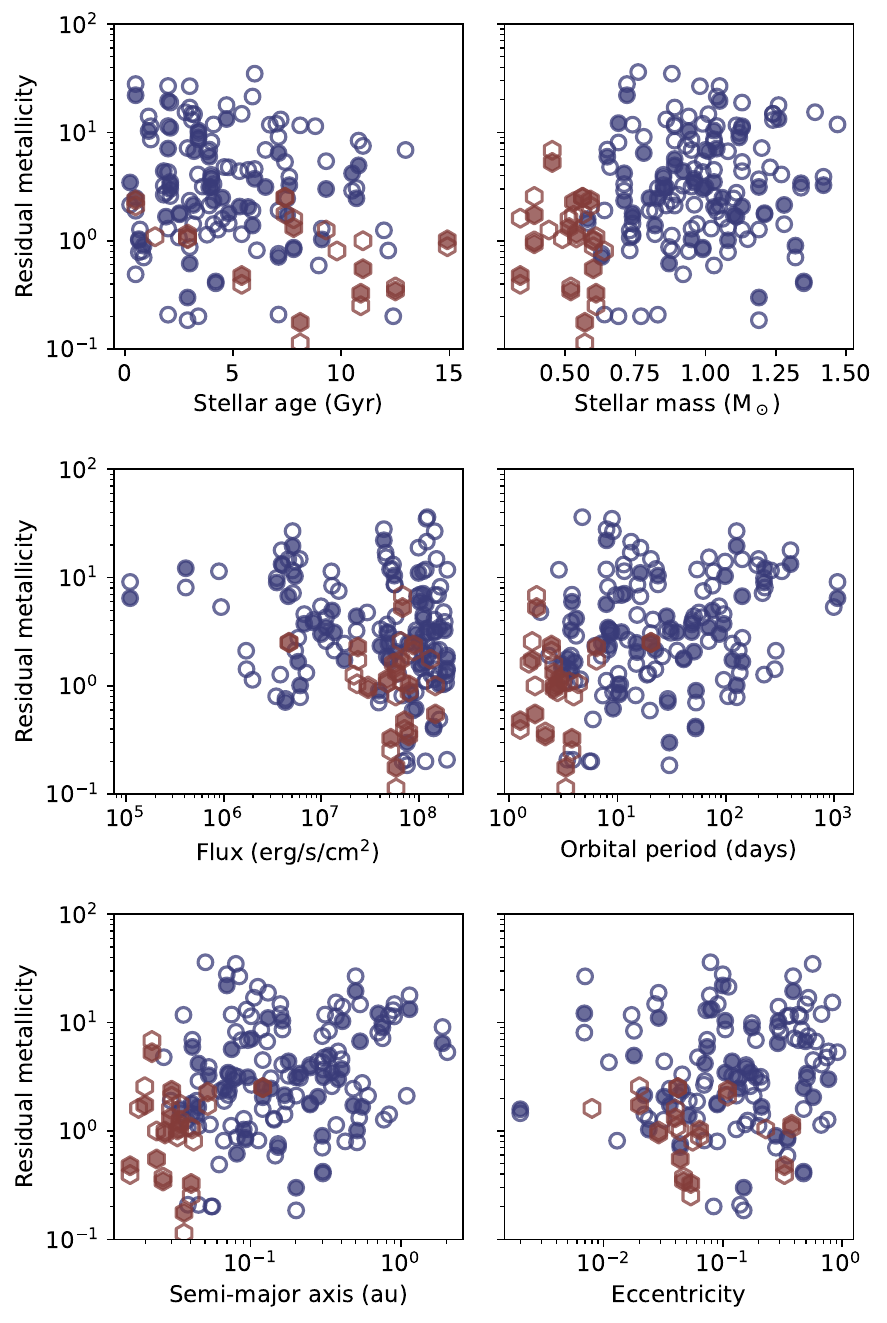}
    \caption{The residual normalised metallicity (the ratio of the calculated and predicted values) against various planetary and stellar properties. Planets around FGK and M-dwarf stars are depicted in blue circles and red hexagons. Filled symbols mark planets in the mass-limited sample (see text for further details).}
    \label{fig:correlations_zbulk_rel_residual}
\end{figure}

\begin{figure}
    \centering
    \includegraphics[width=\columnwidth]{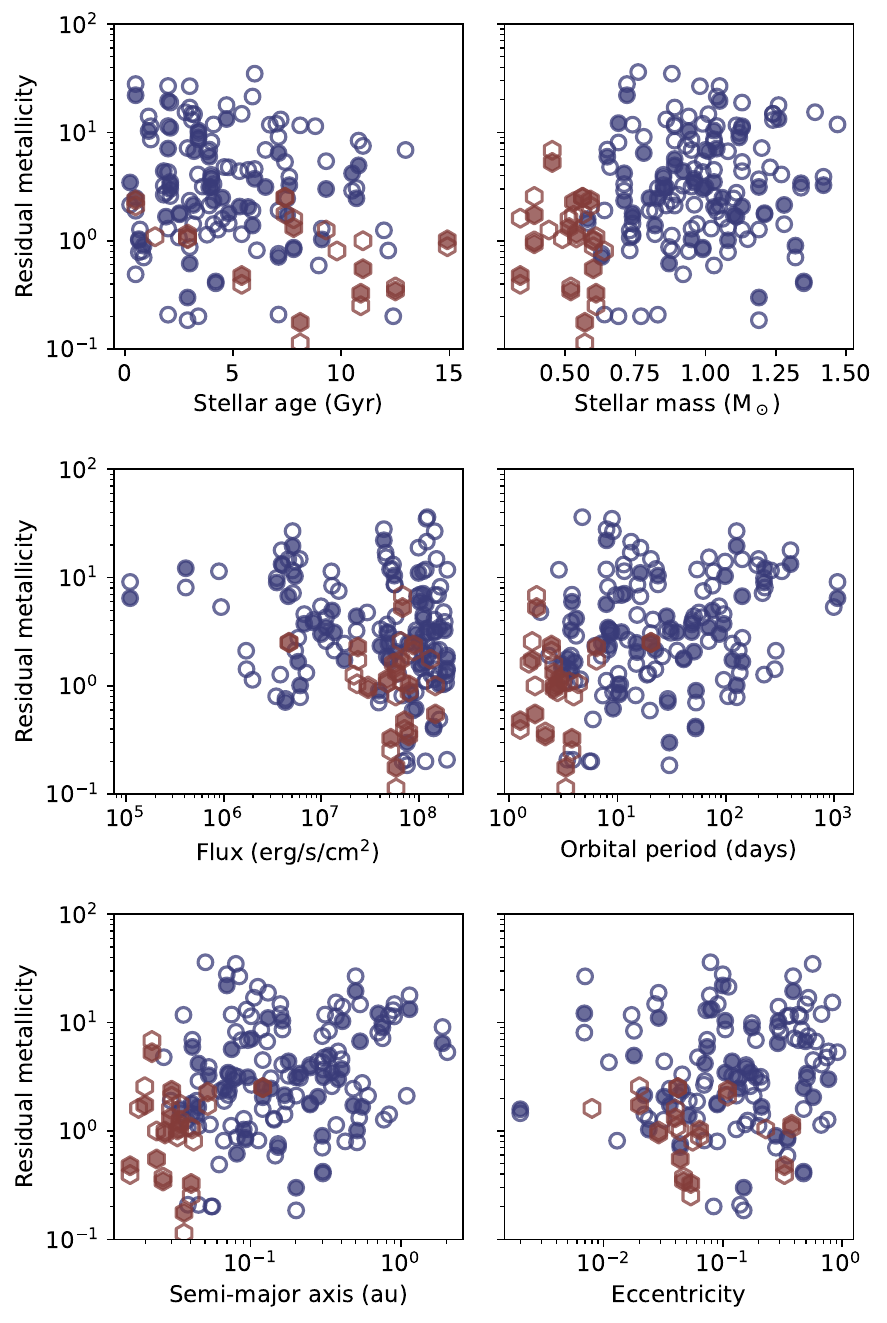}
    \caption{The panels show the residual heavy-element mass (the ratio of the calculated and predicted values) against various planetary and stellar properties. Planets around FGK and M-dwarf stars are depicted in blue circles and red hexagons. Filled symbols mark planets in the mass-limited sample. See text for details.}
    \label{fig:correlations_mbulk_residual}
\end{figure}

\end{document}